\documentclass[12pt,a4paper]{article}

\def\ben{\begin{enumerate}} \def\een{\end{enumerate}}
\def\beq{\begin{equation}} \def\eeq{\end{equation}}
\def\beqn{\begin{equation*}} \def\eeqn{\end{equation*}}
\def\bea{\begin{eqnarray}} \def\eea{\end{eqnarray}}
\def\ba{\begin{array}} \def\ea{\end{array}}
\def\beann{\begin{eqnarray*}} \def\eeann{\end{eqnarray*}}
\def\beasn{\begin{sneqnarray}} \def\eeasn{\end{sneqnarray}}
\def\bi{\begin{itemize}} \def\ei{\end{itemize}}
\def\be{\begin{enumerate}} \def\ee{\end{enumerate}}

\def\o{\"o}
\def\ea{\'e}

\def\mf{\mathfrak}

\usepackage{graphicx}
\usepackage{xcolor}
\usepackage[paper=a4paper,left=30mm,right=30mm,top=30mm,bottom=30mm]{geometry}
\usepackage{amsmath}
\usepackage{amsthm}
\usepackage{amssymb}
\usepackage{eufrak}
\numberwithin{equation}{section}
\title{Observables and Hamilton-Jacobi approaches to general relativity. I. The Earlier History}

\author
{Donald Salisbury$^{1,2}$\\
\\
\normalsize{$^{1}$Austin College, 900 North Grand Ave, Sherman, Texas 75090, USA}\\
\normalsize{$^{2}$Max Planck Institute for the History of Science,}\\
\normalsize{Boltzmannstrasse 22, 14195 Berlin, Germany}}

\date{ }

\begin{document}

\maketitle

\begin{abstract}

The main focus is on the Hamilton-Jacobi techniques in classical general relativity that were pursued by Peter Bergmann and Arthur Komar in the 1960’s and 1970’s. They placed special emphasis on the ability to construct the factor group of canonical transformations, where the four-dimensional diffeomorphism phase space transformations were factored out. Equivalence classes were identified by a set of phase space functions that were invariant under the action of the four-dimensional diffeomorphism group. This is contrasted and compared with approaches of Paul Weiss, Julian Schwinger, Richard Arnowitt, Stanley Deser, Charles Misner, Karel Kuchar – and especially the geometrodynamical program of John Wheeler and Bryce DeWitt where diffeomorphism symmetry is replaced by a notion of multifingered time. The origins of all of these approaches are traced to Elie Cartan’s invariant integral formulation of classical dynamics. A related correspondence concerning the thin sandwich dispute is also documented.
\end{abstract}

\section{Introduction}

Given that the classical Hamilton-Jacobi theory played such a prominent role in the formulation of Schr\o dinger quantum wave mechanics, it is remarkable that general relativists did not make a serious effort in applying the technique until the early 1960's. The origins of this attempt in flat spacetime field theory can be traced to Dirac's 
groundbreaking 1933 paper, {\it The Lagrangian in quantum mechanics} \cite{Dirac:1933ab}, where he established the link, for systems with a finite number of degrees of freedom, between initial and final classical action configuration variables, Hamilton principal functions, and quantum transition amplitudes. The central physical object was the action evaluated on classical solution trajectories. In what I identify in this essay as Hamilton-Jacobi theory I will take a broad view and include under this rubric several approaches in which the boundary terms of the integrated action play a fundamental role. One would wish for these terms to appear as a quantum phase, as did Schr\o dinger, and for this reason, as we shall see, it should apparently refer only to objects which are invariant under whatever local gauge symmetries that are present. It is this demand, in the context of general relativity, that has led historically to the still unresolved conflict between the Syracuse school led by Peter Bergmann, and the Wheeler DeWitt approaches. 

Following a brief historical overview, extending from \'Elie Cartan's introduction of the Poincar\'e-Cartan form,  I  later focus on the work of Bergmann and his collaborators from the 1960's into the early 1970's, where appropriate comparing and contrasting with contemporary work in the `opposing' camps. As far as possible I will try to structure this essay in the chronological order in which problems were addressed by the Syracuse associates - a group with which I identify, having learned my constraint Hamiltonian approach from my thesis advisor, Peter Bergmann.  Considerable progress had already been made by Bergmann and his collaborators in understanding the variation of the gravitational action about physically distinct solutions of Einstein's equations, and the subgroup of invariant variations that corresponded to a change of spacetime coordinates. I will briefly summarize the detailed analysis that appears in \cite{Salisbury:2020aa}. The resulting fundamental ideas concerning the unphysical nature of coordinates never received in Bergmann's lifetime the attention that he felt they deserved. They were the impetus for my own work with several collaborators, and I will turn to this and additional insights in \cite{Salisbury:2021ad}

I begin in Section \ref{weiss} with a capsule historical overview of the use by of Hamilton-Jacobi techniques in quantum field theory by Paul Weiss, with an emphasis on the treatment of gauge invariant observables. Section \ref{schwinger} is devoted to Julian Schwinger's development of these themes in his quantum action principle. In Section \ref{bergmann} I summarize the work of Bergmann and his school that led to his introduction of the classical reduced phase space in which the elements were understood to be invariant under the action of the full four-dimensional diffeomorphism group. I will include the introduction by Arthur Komar of curvature-based intrinsic coordinates. As we shall see in Section 5 there is a crucial difference between these coordinates and those proposed by Arnowitt, Deser, and Misner (ADM). The latter quantities were not spacetime scalars. These authors did not believe it either relevant or necessary to take into account this full symmetry at the classical level.\footnote{Quoting a remark that Deser made to myself and Dean Rickles in an interview that we conducted with him in 2011 regarding general coordinate covariance ``There is a fundamental conflict between the classical formalism and blindly pushing it to the quantum level." Upon which I commented that the position that he had represented historically is that there is no benefit to be gained from having pursued the classical theory of general covariance. He replied that that was indeed the case.}  Karel Kucha\v{r}, who initiated his own related program in 1972, did not believe that this symmetry existed, rather, he promoted a view that had originally been expressed by John Wheeler, that one was simply witnessing the evolution of ``multifingered time'. Nevertheless, as is discussed in \cite{Salisbury:2021ad},  the canonical change of variable approaches in the action that had been promoted by both  ADM and Kucha\v{r} can be profitably employed in the intrinsic Hamilton-Jacobi theory. I address in Section \ref{BKHJ} the Bergmann and Komar Hamilton-Jacobi approach to general relativity. Wheeler's program did famously employ Hamilton-Jacobi reasoning, as laid out in detail in his 1967 Battelle Rencontre  lectures \cite{Wheeler:1968aa}, and summarized in Berlin in 1968 \cite{Wheeler:1968ac}.\footnote{This lecture is, beginning on page 19, a translation into German of the Battelle Rencontres lectures. Dieter Brill assisted with the translation, and as Brill communicated to me in April, 2018, this is consistent with ``his famous cutting and pasting (literally) of other publications".}.  Although Anderson had already in 1955 begun to consider a Hamiltoni-Jacobi approach to generally covariant systems, it is likely that in response to Wheeler's work Bergmann himself in 1966 \cite{Bergmann:1966aa} began to address the applicability of Hamilton-Jacobi analysis.\footnote{Bergmann was however one of the first to present the Hamilton-Jacobi technique in an introductory mechanics text \cite{Bergmann:1949ac}. And in the companion volume he explicitly explored the link to quantum mechanics \cite{Bergmann:1951ab}} Komar had become by this time a close collaborator, and there followed both a joint paper \cite{Bergmann:1970ab}, and several groundbreaking papers by Komar alone \cite{Komar:1967aa,Komar:1968aa,Komar:1970aa,Komar:1971aa}. Ultimately, Komar was able to show how classical solutions of the Hamilton-Jacobi equation could be interpreted as identifying the reduced phase space that results from the quotienting of the full four-dimensional diffeomorphism group. This will be contrasted with the Wheeler-DeWitt approach to quantum gravity as summarized in the first of DeWitt's 1967 trilogy \cite{DeWitt:1967aa}.

As we shall see in Section \ref{sandwich}, Bergmann objected strenuously to Wheeler's disregard of the full diffeomorphism symmetry. A debate  ensued between Bergmann and Wheeler on one of the foundation principles of Wheelers geometrodynamics program. A centerpiece of this program is Wheeler's identification of 3-geometries as legitimate observables in his action-based approach. Consistent with this choice was his insistence that the fixing of initial and final 3-geometries would correspond to the fixation of a unique 4-geometry. This claim in the case of infinitesimal temporal evolution became known as the `thin sandwich conjecture'. Wheeler sought Bergmann's council on the viability of this idea, and I will discuss their correspondence and related published work, informed also by Komar.

\section{Hamilton-Jacobi analysis and quantum field theory} \label{weiss}

I will begin with a brief and therefore incomplete historical account of the invention of the classical Hamilton-Jacobi method and its application in quantum field theory. The principal quantum contributors are Paul Dirac, Max Born, Paul Weiss, and Julian Schwinger. In 1933 Dirac appealed to the Lagrangian formulation of quantum mechanics  \cite{Dirac:1933ab} where he showed first of all that for a quantum system represented by configuration eigenstates $|q'_i>$ of a complete set of commuting operators $\hat q_i$, with $i = 1 \ldots n$, and alternative canonical operators $\hat Q_i$ with eigenstates $|Q'_i>$, with therefore
\beq
<q' |Q'> = e^{i U(q,Q)/\hbar}, \label{qQ}
\eeq
and letting the momenta conjugate to $\hat q_i$ and $\hat Q_i$ be $\hat p^i$ and $\hat P^i$, respectively, then it follows from the assumption that $[\hat q_i, \hat p^j ] = i \hbar \delta_i^j$ that
$$
\hat p^i = \frac{\partial U(\hat q, \hat Q)}{\partial \hat q_i},
$$
and
$$
\hat P^i =- \frac{\partial U(\hat q, \hat Q)}{\partial \hat Q_i}.
$$
Dirac pointed out that these are the quantum analogues of the expressions for the momenta in terms of the classical generating function $U(q,Q)$.\footnote{\cite{Dirac:1933ab}, p. 315, following equation (7), Dirac gave these relations as between ``operators or dynamical variables'' provided the derivatives are ``well-ordered".} 

This immediately suggests a relation to classical Hamilton-Jacobi theory. Rather than recount its origins, I will go directly to a theoretically distinct approach to classical mechanics that was formulated by \'Elie Cartan \cite{Cartan:1922aa}.\footnote{See \cite{Yourgrau:1968aa} for more detail on the historical origins.}  He showed that the classical equations of motion for $n$ configuration variables $q^i$ could be derived from a new invariance principle involving the integral over a family of solution trajectories which assume the values $q^i(t;s)$ and $v^i(t;s)$ at a given time $t$, with the parameter  $s$ taking values between $0$ and $1$, taking $q^i(0)= q^i(1)$.\footnote{Cartan used the letter $\alpha$ rather than $s$.} The variables $v^i(t;s)$ are taken to be independent of $q^i(t;s)$, i.e., it is not assumed that $v^i(t;s) = \frac{dq^i(t;s)}{dt}$.The new principle is the requirement that the closed integral at time $t$ of a quantity he called the `quantity of motion-energy' ({\it quantit\'e  de mouvement}) be independent of whichever closed path one chooses on the tube of trajectories, or in other words, that it be independent of $t$. The invariant integral takes the form\footnote{Cartan dealt explicitly with the standard particle Lagrangian $L = \frac{m}{2}\left((v^1)^2 +v^2)^2 +v^3)^2 \right)- U(v)$.}
\beq
I = \oint \left(\frac{\partial L(q,v)}{\partial v^i} \delta q^i -\left( \frac{\partial L(q,v)}{\partial v^i} v^i - L(q,v)\right)\delta t \right), \label{I}
\eeq
where $\delta q^i := \frac{\partial q^i}{\partial s} ds$ and $\delta t:= \frac{\partial t}{\partial s} ds$. Cartan proved that the invariance property resulted in the familiar classical equations of motion
$$
\frac{d q^i}{dt}= v^i,
$$
and
$$
\frac{\partial  L(q,v)}{\partial q^i} -\frac{d}{dt}\left(\frac{\partial L(q,v)}{\partial v^i} \right) = 0.
$$
Although he did not consider non-holonomic constrained systems, it will be shown in \cite{Salisbury:2021ad}  how his reasoning can be extended to this case and indeed to gauge field theories including general relativity. But most pertinent  for my discussion at this stage is the observation that the invariant integral can be written in terms of phase space variables via the Legendre transformation $p_i = \frac{\partial L(q,v)}{\partial v^i} $ as 
\beq
I = \oint \left( p_i \delta q^i - H(q,p) \delta t \right), \label{I2}
\eeq
where $H(q,p) $ is the Hamiltonian. Cartan thereby extended Poincar\'e's earlier notion of relative integral invariant from 1890 \cite{Poincare:1890aa} by including $-H \delta t$. Poincaré reserved the term `relative' to this case where the loop is closed. The appellation applied also to higher dimensional closed parametric surfaces at a fixed time \cite{Poincare:1899aa}. 

It will be appropriate in laying the foundation for the canonical quantization program to show how the presumed invariance of $I$ leads to the Euler-Lagrange equations and addition to Lagrange and Poisson brackets. As we shall see, Weiss extended the latter results to field theory. A new extension to gauge theories will be presented in \cite{Salisbury:2021ad}. For this purpose I assume, as did Cartan, that $H(q,p) = \frac{1}{2 m} p^2 + V(q)$. The invariant integral at fixed $t$ in this case is
\beq
I = \oint  \left[p_{i} \delta q^i -\left(\frac{1}{2 m} p^2 + V(q) \right)\delta t \right].
\eeq
Following Cartan, we require that independent variations of the variables at time $t$ to result in no change. Representing these changes by $d$ we have
\bea
0 = dI &=& \oint  \left[dp_{i} \delta q^i +p_{i}d \delta q^i -\left(\frac{1}{ m} p_i dp_i + \frac{\partial V(q)}{\partial q^i} dq^i \right) \delta t-\left(\frac{1}{2 m} p^2 + V(q) \right) d\delta t \right] \nonumber \\
&=&\oint \left[dp_{i} \delta q^i - \delta p_i dq^i -\left(\frac{1}{ m} p_i dp_i + \frac{\partial V(q)}{\partial q^i} dq^i \right) \delta t \right.\nonumber \\
&+& \left. \left(\frac{1}{ m} p_i \delta p_i + \frac{\partial V(q)}{\partial q^i} \delta q^i \right)\right] dt +\oint \delta \left(p_i dq^i -\left(\frac{1}{2 m} p^2 + V(q) \right)dt \right) 
\eea
The final integral vanishes by assumption. Now setting the coefficients of each $d$ increment equal to zero, we deduce that $p^i = m \frac{dq^i}{dt}$ and $\frac{d p_i}{dt} = - \frac{\partial V}{\partial q^i}$, as desired.

This is also the route that Cartan took to derive Lagrange and Poisson brackets, as I now show. The idea is to fill the `tube' of solutions parameterized by $s$, labeling them by the parameter set $\omega_A$ for $A=1,2$. Then one computes the line integral of $p_i \delta q^i - H(q,p) \delta t $ around the infinitesimal closed path $(\omega_1,\omega_2) \rightarrow (\omega_1+ d\omega_1,\omega_2) \rightarrow (\omega_1+ d\omega_1,\omega_2+ d\omega_2)  \rightarrow  (\omega_1+ d\omega_1,\omega_2) \rightarrow (\omega_1,\omega_2)$. The result is (Stoke's theorem) 
\beq
2 \left(\frac{\partial p_i}{\partial \omega_{[A}} \frac{\partial q^i}{\partial \omega_{B]}} +  \frac{\partial H}{\partial \omega_{[A}} \frac{\partial t}{\partial \omega_{B]}} \right) d\omega_A d\omega_B.
\eeq
The quantity enclosed in the parentheses is therefore independent of $t$. With its inclusion of $H$ and $t$ it is a generalization of  Poincar\'e's absolute integral invariant. The first term is the invariant Lagrange bracket,
\beq
\left[ \omega_1, \omega_2 \right] :=2  \frac{\partial p_i}{\partial \omega_{[1}} \frac{\partial q^i}{\partial \omega_{2]}}.
\eeq
It's inverse is the Poisson bracket,
\beq
\left\{\omega_1, \omega_2\right\}: =  \frac{\partial \omega_1}{\partial q^i} \frac{\partial \omega_2}{\partial p_i}-\frac{\partial \omega_1}{\partial p_i} \frac{\partial \omega_2}{\partial q^i}
\eeq

As Cartan observes in 1922, the  integral  invariant (\ref{I2}) offers the quickest route to Hamilton-Jacobi theory. The expression $p_i dq^i$ is now known as the Poincar\'e one form, and with the extension to include $-H(q,p) \delta t $, it is called the Poincar\'e-Cartan one form. Cartan noted  that the closed integral over $s$ of the latter form would be unaltered by the addition of an exact differential $\delta G$. Citing Jacobi \cite{Jacobi:1837aa}, Cartan observed\footnote{\cite{Cartan:1922aa}, p. 15} this this implied that one could carry out a change of canonical variables $(q^i,p_j)$ to $(Q^i,P_j)$ which did not change the value of the integral (\ref{I}) provided
\beq
p_i \delta q^i - P_j \delta Q^J - \left(H(q,p) - K(Q,P)\right)\delta t = \delta G(q,Q,t). \label{CHJ1}
\eeq
Finally, if one were to choose a change of variables such that $K$ were zero, resulting in new constant phase space coordinates $\left(\beta_i, \alpha^i\right)$, then letting $S$ represent the corresponding generating function, (\ref{CHJ1}) results in
\beq
p_i \delta q^i - P_j \delta Q^J - H(q,p) \delta t = \frac{\partial S}{\partial q^i} \delta q^i + \frac{\partial S}{\partial \alpha^j} \delta Q^j +  \frac{\partial S}{\partial t} \delta t, \label{CHJ2}
\eeq
from which we deduce the Hamilton-Jacobi equations
\beq
p_i = \frac{\partial S}{\partial q^i},
\eeq
\beq
\frac{\partial S}{\partial t} + H\left(q, \frac{\partial S}{\partial q^i}\right) = 0, \label{CHJ3}
\eeq
and
\beq
\beta^j = - \frac{\partial S}{\partial \alpha^j}.
\eeq

We are invited to consider the complete solution of (\ref{CHJ3}) as depending on $n$ constants $\alpha^i$,  We conclude that $S(q,t; \alpha)$ may be interpreted as transforming the canonical coordinates at time $t$ to initial conditions at time zero! 

The relation of Cartan's new invariance principle to that of Hamilton is easily obtained, and the corresponding variations will play a central role in this paper. Consider a variation of the action 
\beq
S = \int_{t_0}^{t_1} L(q, \dot q) dt,
\eeq
where $\dot q^i := \frac{dq^i}{dt}$, where we again assume that we vary from solution to neighboring solutions, parameterized by $s$, as well as vary the time, so the net variation in $q^i$ is 
\beq
\delta q^i := \frac{\partial q^i}{\partial s} \delta s + \dot q^i \frac{dt}{ds} \delta s =: \delta_0 q^i  + \dot q^i \delta t. \label{deltaq}
\eeq
So the variation of $q^i$ at fixed $t$ is 
\beq
\delta_0 q^i = \delta q^i -  \dot q^i \delta t. \label{delta0q}
\eeq
 Therefore the net variation of the action is
\bea
\delta S &=& \int_{t_0}^{t_1}  \left(\frac{\partial L}{\partial q^i} \delta_0 q^i+ \frac{\partial L}{\partial \dot q^i} \frac{d}{dt} \delta_0 q^i  \right) + \left.L \delta t\right|_{t_0}^{t_1} \nonumber \\
&=&  \int_{t_0}^{t_1} \left(\frac{\partial L}{\partial q^i} - \frac{d}{dt}\left( \frac{\partial L}{\partial \dot q^i} \right)\right) \delta_0 q^i + \left.\left[ \frac{\partial L}{\partial \dot q^i} \left(\delta q^i -  \dot q^i \delta t \right)-L \delta t \right]\right|_{t_0}^{t_1}, \nonumber \\
&=&  \left.\left[ \frac{\partial L}{\partial \dot q^i} \delta q^i - \left( \frac{\partial L}{\partial \dot q^i}\dot q^i -L\right)\delta t \right]\right|_{t_0}^{t_1} \label{deltaSC}
\eea
where in the first line I took into account the change in the range of integration, and in the second line I integrated by parts. The integral vanishes by assumption - and we note that the differential evaluated at the endpoints of integration is precisely the integrand that appears in Cartan's new invariance principle. Indeed, as noted by Cartan, if we were to let the variations at time $t_0$ vanish and if we let $\delta q^i(t_1) = \dot q^i(t_1) \delta t(t_1)$, then the resultant variation in $S$ is simply $L(t_1) \delta t_1$. This same result can, of course, be obtained from (\ref{deltaSC}) in terms of the position-velocity functions, again taking the variations at the initial time to vanish, i.e.
\beq
\delta S =  \frac{\partial L}{\partial \dot q^i} \delta q^i - \left( \frac{\partial L}{\partial \dot q^i}\dot q^i -L\right)\delta t.
\eeq
When the Lagrangian is non-singular, so that the velocities can be expressed in terms of the momenta, then
\beq
\delta S = p_i \delta q^i - H(q,p) \delta t. \label{dSns}
\eeq
In the singular case constraining relations will arise but as we shall see they will make vanishing contributions to the action increment. Note that (\ref{dSns}) also follows from (\ref{CHJ2}), taking $\delta Q^i = \delta \alpha^i = 0$, in which case we confirm that $\delta S$ is indeed the variation of the action.

Note also that Cartan's new invariance principle follows almost directly from (\ref{deltaSC}); the integrals around closed loops of solutions must be independent of the times $t_0$ and $t_1$. (I say 'almost' since Cartan works with supposedly independent variables $q^i$ and $v^j$. The more general case is treated in \cite{Salisbury:2021ad}.)

It is noteworthy that Cartan already suggested at this time that his integrand could be altered to place the positions $q_i$ and time $t$ on an equal footing - as would presumably be required of relativistic dynamics - by letting the time $t$ itself depend on an independent variable $\tau$, i.e. $t = q_0(\tau)$ with corresponding momentum $-H$. He in fact wrote done the resulting generally covariant Lagrangian for particle motion in general relativity.\footnote{\cite{Cartan:1922aa}, p. 14}  

It is also pertinent to note here that Cartan's procedure constituted the basis of what became known as the finite-dimensional multisymplectic approach to field theory in which the spacetime coordinates and the finite number of field components constitute the formal basis. I will explain this further in the following discussion of Weiss, and later in discussing the geometric differential theory in \cite{Salisbury:2021ad}.

As we have seen, Poincaré's relative integral invariant plays a special role in canonical quantization programs since it can be employed to derive Laplace brackets and, for non-singular models, their inverses which are Poisson brackets. This was Paul Weiss's motivation for undertaking his study. He expected that these considerations could lead to Poisson brackets in field theory which could then be promoted to a quantum commutator algebra.

I return now to Dirac's second conclusion although he did not give the following reasoning in his paper. The conclusion is a natural extension of his first observation. One can interpret the transition in time as a canonical transformation generated by $S(q,t; \alpha)$, so that, corresponding to (\ref{qQ}),
\beq
\left<q_{t_2} | q_{t_1} \right> = e^{i S(q_{t_2}, t_2-t_1; q_{t_1})/\hbar}.
\eeq
Dirac did point out in this paper that this relation could be easily generalized to apply to quantum field theory. But the first to publish this generalization was Paul Weiss.

Before addressing the generalizations by Dirac and Schwinger, it is appropriate here to briefly review Schr\o dinger's use of Hamilton-Jacobi theory in the reasoning that led to his quantum mechanical wave equation.\footnote{Schr\o dinger discussed the link of his wave equation to classical Hamilton-Jacobi theory in his second article in 1926 \cite{Schrodinger:1926ab}, but this classical mechanical origin is clear from his unpublished notebooks. See \cite{Joas:2009aa}  for a discussion.} A power series expansion in ascending powers of $ \left(\frac{\hbar}{i}\right)$ of the Hamilton principal function $S$ was introduced independently by Wentzel \cite{Wentzel:1926aa} and Brillouin \cite{Brillouin:1926aa} in 1926, $S = S_0 + \left(\frac{\hbar}{i} \right)S_1 +  \left(\frac{\hbar}{i} \right)^2 S_2 + \ldots$. They in turn with Kramers \cite{Kramers:1926aa} in 1926 established what became known as the WKB approximation. Write the single particle Schr\o dinger equation with Hamiltonian $H(q,p)$ as $H\left(q,-i \hbar\frac{\partial }{\partial q}\right) \Psi = i \hbar \frac{\partial \Psi}{\partial t} $ and assume that $\Psi =  {\cal A} \exp \left(i S/\hbar\right)$ and that $\left|\frac{\partial {\cal A}}{\partial q}\right|<< \left| {\cal A}\frac{\partial S}{\partial q}\right|$. Then under these assumptions the first order term is precisely the Hamilton-Jacobi equation for (\ref{CHJ2}) for $S_1$.

I turn now to Paul Weiss \cite{Weiss:1938aa,Weiss:1938ab}. The inspiration for his work was the foundational invariant integral presented by Cartan, and also, as we shall see, pursued by Theophile DeDonder. 
He in fact sought a generalization to field theory.
Weiss was a relatively unknown student of Max Born and Paul Dirac.\footnote{See \cite{Rickles:2015aa} and \cite{Rickles:2020aa} for biographical sketches and a critical evaluations of his work.}  Dirac was later on clearly informed by Weiss' field-theoretic Hamilton-Jacobi approach, and in fact served as Weiss' Ph. D. supervisor at Cambridge University after his former advisor Max Born left Cambridge to assume the Tait Chair of Natural Philosophy at Edinburgh. Schwinger explicitly noted his debt to Dirac in the formulation of the Schwinger quantum action principle though I have not been able to find any explicit reference to Weiss in Schwinger's work. Schwinger did refer to Dirac's  Quantum Theory of Localizable Dynamical Systems \cite{Dirac:1948aa} in his 1948 paper on a covariant formulation of quantum electrodynamics \cite{Schwinger:1948aa}. And in this cited paper Dirac made extensive reference to Weiss' work. 

Weiss extended the Hamiltonian-Jacobi variational approach to quantum special-relativistic field theory. In so doing, in addition to citing both Cartan \cite{Cartan:1922aa} and de Donder \cite{Donder:1935aa}, he also referred to Born \cite{Born:1934ab} and his collaboration with Infeld \cite{Born:1934aa}.  Weiss  employed quantum operators and was the first to make use of arbitrary curved spacelike foliations (in flat spacetime). He did so as a means of exhibiting the Lorentz covariance of the theory.  This covariance follows from his result that the dynamics takes the same form for arbitrary orientations of spacelike plane foliations in a given Lorentz frame. His initial paper appeared in 1936, \cite{Weiss:1936aa} and he followed up in 1938 with a demonstration of the equivalence of his quantum procedure to the canonical approach in which classical Poisson brackets are replaced by quantum commutators \cite{Weiss:1938aa,Weiss:1938ab}.  

Weiss demonstrated in 1938  the equivalence of the Hamilton-Jacobi-based quantum theory with the canonical quantization approach. He considered an action of the form\footnote{I have translated into more familiar notation, closely related to Schwinger's own in his original 1951 presentation of his quantum action principle \cite{Schwinger:1951aa}. Weiss represented his generic fields by $z$ rather than $\phi$. Also, initially Weiss conceived of the fields $\phi_\alpha$ as operators, but then for the purposes of the 1938 paper reverted to the classical theory so as to recover Poisson brackets.}
\beq
S = \int_{\sigma_1}^{\sigma_2} {\cal L}\left(\phi_\alpha,  \phi_{\alpha, \mu}  \right) d^4 x. \label{SW}
\eeq
The integral extends from the initial space like surface $\sigma_1$ to the final space like surface $\sigma_2$ . Similarly to the finite-dimensional case treated by Cartan, Weiss defined a ``complete variation" in which both the fields $\phi_\alpha(x)$ and the spacetime coordinates $x^\mu$ are varied, and we suppose, as did Cartan, that these variations lead to new solutions of the dynamical equations which we can suppose are labeled by a parameter $s$. We therefore generalize the net variation (\ref{deltaq}) to
\beq
\delta \phi_\alpha := \frac{\partial \phi_\alpha}{\partial s} \delta s + \frac{\partial \phi_\alpha}{\partial x^\mu} \frac{d x^\mu}{ds} \delta s =: \delta_0 \phi_\alpha+\phi_{\alpha, \mu} \delta x^\mu,
\eeq
and  the variation (\ref{delta0q}) at fixed spacetime coordinate $x^\mu$ becomes
\beq
\delta_0 \phi_\alpha (x) :=\delta   \phi_\alpha (x) -  \phi_{\alpha, \mu} (x ) \delta x^\mu (x ). \label{delta0}
\eeq
Weiss viewed the variations $\delta x^\mu$ and $\delta \phi_\alpha$  as fundamental because they represented the displacement in $\left(x^\mu, \phi_\alpha \right)$ space.\footnote{\cite{Weiss:1938aa}, p. 105. }

It is important to note that it will be understood in the analysis below that each of the varied $\phi_\alpha$, namely $\phi_\alpha + \delta_0 \phi_\alpha$ and $\phi_\alpha + \delta \phi_\alpha$ must also be solutions of the Euler-Lagrange equations that follow from the variation of ${\cal L}$ since we will be interpreting the varied action as a function of the initial and final values of the action when evaluated on solutions. 

Using $d^4 x' = \left|\frac{\partial x'}{\partial x}  \right| = \left(1 + \frac{\partial \delta^\mu(x)}{\partial x^\mu}  \right)$ he found that the varied action is\footnote{He cited \cite{Cartan:1922aa} for proof, stating that he had learned this method from Born in Cambridge in 1933.}
\beq
\delta S = \int_{\sigma_1}^{\sigma_2} \frac{\delta {\cal L}}{\delta \phi_\alpha} \delta_0 \phi_\alpha d^4 x +  \int\left(\frac{\partial {\cal L}}{\partial \phi_{\alpha, \mu}}\delta_0 \phi_\alpha +  {\cal L} \delta x^\mu \right) d\Sigma_\mu \left. \right|_{\sigma_1}^{\sigma_2}. 
\eeq
Weiss employed a parameterization which I will represent as $x^\mu(u^0, \vec u )$ such that the boundary hypersurfaces correspond to $u^0= u^0_1$ and $u^0= u^0_2$.\footnote{Weiss used the notation $w$ rather than $u^0$} Weiss calls this a ``natural coordinate system", with the additional condition that $u^0$ increase in the direction of the normal to the $\vec u = constant$ surface. The surface area element is
\beq
d\Sigma_\mu =\frac{1}{3 !} \epsilon_{\mu \nu \rho \sigma} \frac{\partial x^\nu}{\partial u^1}\frac{\partial x^\rho}{\partial u^2} \frac{\partial x^\sigma}{\partial u^3} d^3 u =: n_\mu d^3 u. 
\eeq
Furthermore, he assumed that variations vanished at spatial infinity.
Thus when the Euler-Lagrange equations $\frac{\delta {\cal L}}{\delta \phi_\alpha} = 0$ are satisfied, we have\footnote{I have altered Weiss' notation to conform to modern usage. He represented my $\pi^\alpha$ by $P_\alpha$ and ${\cal P}_\mu$ by $X_\mu$.}
\bea
\delta S &=& \int \left[ \pi^{\alpha \mu} \delta \phi_\alpha - \left(\pi^{\alpha \mu} \phi_{\alpha, \nu} -{\cal L} 
\delta^\mu_\nu \right)\delta x^\nu  \right] n_\mu d^3 u \left. \right|_{\sigma_1}^{\sigma_2} \label{dS1a}\\
&=&  \int \left[ \pi^\alpha  \delta \phi_\alpha - {\cal P}_\nu \delta x^\nu  \right]  d^3 u \left. \right|_{\sigma_1}^{\sigma_2} \label{dS2}
\eea
where
\beq
\pi^{\alpha \mu}  := \frac{\partial {\cal L}}{\partial \phi_{\alpha, \mu}},
\eeq
\beq
\pi^\alpha := \pi^{\alpha \mu} n_\mu,
\eeq
and
\beq
{\cal P}_\mu := \pi^\alpha \phi_{\alpha, \mu} - {\cal L} n_\mu. \label{calpmu}
\eeq

Although Weiss did not write down the corresponding Hamilton-Jacobi equation {\it per se}, it does follow from (\ref{dS2}), as in 
Cartan's derivation, as a consequence of
\beq
dS = \pi^\alpha d\phi_\alpha - {\cal P}_\mu dx^\mu, \label{WdS}
\eeq
that
\beq
 \pi^\alpha = \frac{\partial S}{\partial \phi_\alpha}, \label{WHJ1}
\eeq
and
\beq
{\cal P}_\mu = -\frac{\partial S}{\partial x^\mu}. \label{WHJ2}
\eeq

(I note in passing that it is at this stage that Weiss parted company with a `generalized Hamilton-Jacobi procedure' that had been pursued by Weyl \cite{Weyl:1934aa,Weyl:1935aa}, DeDonder \cite{Donder:1935aa} and Carath\'eodory \cite{Caratheogory:1935aa}. They put equal stress on all components of $\pi^{\alpha \mu}$. As a preparation for the later multisymplectic formalism that will be briefly discussed in \cite{Salisbury:2021ad} I will summarize the relevant work in the Appendix.)

Expressed in terms of the natural coordinate system, the boundary variation (\ref{dS2}) becomes
\beq
\delta S =  \int \left[ \pi^\alpha  \delta \phi_\alpha - {\cal H} \delta u^0 - {\cal P}_a  \delta u^a  \right]  d^3 u \left. \right|_{\sigma_1}^{\sigma_2}, \label{dSw}
\eeq
where
\beq
{\cal H} := \pi^\alpha \frac{\partial \phi_\alpha}{\partial u^0} - {\cal L},
\eeq
and
\beq
{\cal P}_a :=- \pi^\alpha \phi_{\alpha, a} \label{pa}
\eeq
In terms of the Legendre transformed variables $\pi^\alpha$ and $\phi_\beta$ the Euler-Lagrange equations become
\beq
\frac{\partial \pi^\alpha}{\partial u^0} = -\frac{\delta H}{\delta \phi_\alpha}, \label{pidot}
\eeq
and
\beq
\frac{\partial \phi_\alpha}{\partial u^0} = \frac{\delta H}{\delta \pi^\alpha}, \label{psidot}
\eeq
where
\beq
H := \int d^3\!u {\cal H}.
\eeq

Weiss' next step was to extend the Poincar\'e Cartan notion of relative integral invariant (\ref{I}) to field theory. This turns out to be a essentially a straightforward extension from a finite index set $i$ to a continuous set parameterized by the $u$. Consider a closed family of solutions parameterized by $\lambda$, such that $\phi_\alpha(u,w,\lambda=0) = \phi_\alpha(u,\lambda=1) $ and $x^\mu(u,\lambda=0) = x^\mu(u,\lambda=1) $. It follows as an immediate generalization of (\ref{I}) that there is a relative integral invariant
$$
I_{field} = \int_0^1 d\lambda \left(  \int \left[ \pi^\alpha  \frac{d \phi_\alpha}{d\lambda} - {\cal P}_\mu \frac{d x^\mu}{d \lambda}   \right]  d^3 u \right).
$$
Then, as had been done in the finite case, Weiss derived from this expression an absolute invariant by filling the one-dimensional loop of trajectories with a two-dimensional set $\omega_1$ and $\omega_2$ and via Stoke's theorem converting the loop intergral to a surface integral. Furthermore, as with Cartan, one can keep the $x^\mu$ fixed. And then, since the parameter area is arbitrary, one obtains the absolute invariant integrand
$$
\int d^3 u \left( \frac{\partial \pi^\alpha}{\partial \omega_1}  \frac{\partial \phi_\alpha}{\partial \omega_2} - \frac{\partial \pi^\alpha}{\partial \omega_2}  \frac{\partial \phi_\alpha}{\partial \omega_1} \right)
$$
which is a generalization to field theory of the Lagrange bracket of $\omega_1$ and $\omega_2$. Finally, suppose that there exists a number $2 \nu$ of parameters $\omega_a(u) $ in terms of which the set $(\phi_\alpha(u), \pi^\beta(u)$ are fixed. Then we can invert the Lagrange bracket to obtain the absolute invariant Poisson bracket $\left\{\omega_a(u), \omega_b(v) \right\}$. 

As we shall see in section \ref{schwinger}, Schwinger's quantum action approach actually receives its prior justification from Weiss' work. But before I  discuss Schwinger I will look briefly at Weiss's proposal for the treatment of gauge invariance in quantum electrodynamics. In applying the preceding analysis to a quantum Hamilton-Jacobi model of electrodynamics, Weiss proposed a means for dealing with the underlying $U(1)$ gauge invariance of electrodynamics \cite{Weiss:1938ab}. This is of special interest  since we find here the first introduction of gauge invariants in the Hamilton-Jacobi formalism. The problems attendant to the extension of the Hamilton-Jacobi method to general relativity with its underlying general covariance symmetry constitute the main focus of the remainder of this paper. 

Weiss applied his technique to free Maxwellian electromagnetism with the Lagrangian
\beq
L_{em} =  -\frac{1}{4} F^{\mu \nu} F_{\mu \nu}, \label{lem}
\eeq
where $F_{\mu \nu}  := \partial_\mu A_\nu - \partial_\nu A_\mu$ is the electromagnetic tensor and $A_\mu = ( -V/c, \vec A)$ the electromagnetic 4-potential. Weiss let his initial value surface be simply $x^0 = constant$, Then the momentum conjugate to $A_0$ is
\beq
p^0 = \frac{\partial L_{em}}{\partial \dot A_0} = 0.
\eeq
This is in modern terminology a primary constraint, and its appearance is a reflection of the fact that that the Legendre matrix $\frac{\partial^2 L_{em}}{\partial \dot A_\mu \partial \dot A_\nu}$ is singular and that therefore it is not possible to solve for all of the $\dot A_\mu$ in terms of the conjugate momenta.\footnote{Strangely, Weiss did not refer to L\'eon Rosenfeld's 1930 opus \cite{Rosenfeld:1930aa}\cite{Rosenfeld:2017ab} in which he presented a constrained Hamiltonian analysis of not only electromagnetic fields in interaction with evolving charged spinorial sources, but also with evolving curved metrics. For a discussion of the application of Rosenfeld's methods to quantum electrodynamics in flat spacetime and Dirac's awareness of his formalism, see \cite{Salisbury:2009ab}. See  \cite{Salisbury:2017aa} for a detailed analysis of Rosenfeld's groundbreaking work in constrained Hamiltonian dynamics. Rosenfeld followed up with an overview of the formalism in his 1932 review of quantum electrodynamics \cite{Rosenfeld:1932aa}. Born cited this work in his initial 1934 Hamilton-Jacobi approach to quantum electrodynamics \cite{Born:1934ab}. As Weiss was Born's student at this time it is hard to imagine that Born would not have brought Rosenfeld's work to Weiss' attention. Even more mysterious is Dirac's evident failure to notify Weiss after Dirac took over Weiss' thesis direction.}
The remaining momenta are
\beq
p^a = \frac{\partial L_{em}}{\partial \dot A_a} = -\frac{1}{c}F^{0a}   =\frac{1}{c^2} \left(\dot A^a + V_{,a} \right)=-\frac{1}{c^2} E^a, \label{pa}
\eeq
where $E^a$ is the electric field. Weiss' Hamiltonian would therefore be
\beq
{\cal H} = \frac{1}{2}\left( -c^2 \vec p^2 + \vec B^2 \right)  +  p^0 \dot A_0 +  p^a \dot A_a,
\eeq
Weiss found this expression unacceptable due to the presence of the term $p^a \dot A_a$ that is not invariant under the gauge transformation $A_\mu \rightarrow A_\mu + \Lambda_{,\mu}$, where $\Lambda$ is an arbitrary function of $x$. He attributed the presence of this term to the fact that the quantities that were being varied in the action were themselves not gauge invariant. This concern is indeed legitimate, though Weiss' procedure for introducing a Hamilton principal function depending only on invariant variables is open to interpretation. A correct interpretation follows from the employment of gauge invariant functionals - which in practice amounts to imposing gauge conditions as we shall also see below in our discussion of gravity. Weiss argued that the variation of the action should depend on the gauge invariant $\dot A_a - c A_{0,a}$, in which case the resulting Hamilton density would be ${\cal H} = \frac{1}{2}\left( c^2 \vec p^2 + \vec B^2 \right)  +  p^0 \dot A_0$. Rosenfeld had however already observed in 1930 that the correct gauge symmetry-based approach was to substitute the particular solution (\ref{pa}) into the Hamiltonian, resulting in
\beq
{\cal H} = \frac{1}{2}\left( c^2 \vec p^2 + \vec B^2 \right)  +  c p^a \dot A_{0,a}+ \lambda p^0,  \label{hem}
\eeq
where $\lambda$ is an arbitrary spacetime function. This procedure has the remarkable dividend that the preservation of the primary constraint $p^0 = 0$ over time requires the secondary constraint $p^a_{,a}=0$. Rosenfeld was indeed the first to present a systematic algorithm for producing secondary and higher order constraints. Note that in taking the secondary constraint into account every term in (\ref{hem}) is gauge invariant. There is in fact  a valid Cartan inspired invariant integral approach to electrodynamics that is fully consistent with the Rosenfeld Bergmann Dirac approach to constrained Hamiltonian dynamics, as will be shown in \cite{Salisbury:2021ad}.

\section{Schwinger's quantum action principle} \label{schwinger}

Classical Hamilton-Jacobi analysis has of course played an important role in the development of quantum mechanics. Perhaps less well known than Schr\"odinger's use of the action in his invention of quantum wave mechanics is the appearance of a quantum action principle in the earliest stages of development of quantum matrix mechanics. Heisenberg published in 1925 his groundbreaking introduction of non-commuting quantum variables \cite{Heisenberg:1925aa}. Inspired by this paper, Born and Jordan shortly thereafter revealed that the quantum matrices obeyed the commutator algebra $[q,p] = i\hbar$  and they postulated a matrix (operator) quantum action principle based on a Lagrangian of the form $p \dot q - H(p,q)$. Jordan continued to pursue the quantum connection to Hamilton's principal function in 1926 , but as we have seen it was Dirac in 1933 \cite{Dirac:1933ab} who according to Schwinger himself most directly influenced Schwinger's quantum field-theoretic action principle.\footnote{"During the 25 year period of quantum electrodynamical development,
there was great formal progress in the manner of presenting the laws of quantum
mechanics, all of which had its inspiration in a paper of Dirac. This paper (which is
No. 26 in the collection, Selected Papers on Quantum Electrodynamics, Dover, 1958)
discussed for the first time the significance of the Lagrangian in quantum mechanics.", \cite{Schwinger:1973aa}, p. 420. Reference is to \cite{Schwinger:1958aa}}

Schwinger proposed that one sever the explicit link to classical field theory by positing a quantum action principle. If this were done properly one could then anticipate recovering the classical theory in an appropriate correspondence limit. Hamilton-Jacobi theory suggested to him the desired quantum dynamical principle. There are three aspects of Schwinger's formulation that are of special relevance in this brief historical account. The first has to do with Dirac's inspiration, and apparently through him, foundations established by Weiss. The second is Schwinger's recognition that the quantum dynamical principle follows strictly only for gauge invariants. The third is Schwinger's use of actions linear in velocity variables.
In my discussion of Weiss in section \ref{weiss} I purposefully employed Schwinger's 1951 notation so as to demonstrate Schwinger's (unacknowledged) indebtedness to Weiss. 

Schwinger defines an action operator  equivalent of (\ref{SW}). The quantum action principle is the postulate that the variation of the action operator depends only on  independent variations $\delta \phi$ of the operators on the surfaces $\sigma_1$ and $\sigma_2$, and on deformations $\delta x^\mu$ of the surfaces. I will refer to Schwinger's 1951 article to show how  the variation of the action appears as a matrix operator expression. I will take our spacelike hypersurfaces to be Lorentz time coordinate $x^0 = c t = constant$ surfaces.

Let us return to Weiss's variation of the action $S = \int_{t_2}^{t_1} dt d^3\!x \left( \tilde {\cal L}\left(\phi_\alpha,  \phi_{\alpha, \mu}  \right) \right)$,\footnote{Schwinger represents the action by $W$ rather than $S$.} where we now conceive of the $\phi_\alpha(\vec x)$ as operators. I let $\tilde {\cal L}$ signify that the Lagrangian is conceived as a function of the fields and field space and time derivatives. I will however, as does Schwinger,  conceive of the variations of $\phi$ as c-numbers, and I will represent c-numbers with a ``prime". Later we consider specific matrix representations of the operators. The varied action is
\bea
\delta S &=&   \int^{t_1}_{t_2}  dt d^3\!x \left(\frac{\partial \tilde {\cal L}}{\partial \phi_\alpha} -\partial_\mu \frac{\partial \tilde {\cal L}}{\partial \phi_{\alpha,\mu}} \right)\delta_0 \phi_\alpha \nonumber \\
&+& \int \left[ \tilde \pi^\alpha  \delta \phi_\alpha - \tilde {\cal H} \delta t - \tilde {\cal P}_a \delta x^a  \right]  d^3 x \left. \right|^{t_1}_{t_2}  \label{deltaS}
\eea
where
\beq
\tilde \pi^\alpha (\phi_\alpha,  \phi_{\alpha, \mu}) := \frac{\partial \tilde {\cal L}}{\partial \dot \phi_\alpha}, \label{pitilde}
\eeq
\beq
\tilde {\cal H} (\phi_\alpha,  \phi_{\alpha, \mu}):= \tilde \pi^\alpha \dot \phi_\alpha - \tilde {\cal L},
\eeq
and
\beq
\tilde {\cal P}_a (\phi_\alpha,  \phi_{\alpha, \mu}):= \tilde \pi^\alpha \phi_{\alpha,a}. 
\eeq

In the classical approach the vanishing of the integral expression in (\ref{deltaS}) for arbitrary variations $\delta_0$ would result in the Euler-Lagrange equations
\beq
\frac{\partial \tilde {\cal L}}{\partial \phi_\alpha} -\partial_\mu \frac{\partial \tilde {\cal L}}{\partial \phi_{\alpha,\mu}} = 0. \label{el}
\eeq

We now assume that we are dealing with a non-singular Lagrangian density that will possess the property that the defining equations for the momenta (\ref{pitilde}) can be solved for the velocities as functions of $\phi$ and $ \pi$,
\beq
\dot \phi_\alpha = \dot \phi_\alpha[\phi, \pi].
\eeq 
We will represent the resulting functionals of $\phi$ and $\pi$ without the ``tilde". Thus we have
\beq
{\cal L}[\phi, \pi] = \tilde {\cal L}\left[\phi, \dot \phi_\alpha[\phi,  \pi]\right], \label{calL}
\eeq
\beq
{\cal H}(\phi, \pi) = \pi^\alpha \dot \phi_\alpha[\phi, \pi] - {\cal L}[\phi, \pi]
\eeq
and
\beq
{\cal P}_a [\phi,  \pi] = \pi^\alpha \phi_{\alpha,a}.
\eeq

Let us immediately rewrite the Euler-Lagrange equations in terms of $\phi$ and $\tilde \pi$. We have already noted in section \ref{weiss} that when the Lagrangian is non-singular we can use the definition of canonical momentum to express the velocities in terms of partial derivatives of the Hamiltonian. We demonstrate this now in an equivalent manner in which the non-singularity is used explicitly. We calculate
\beq
\tilde \pi^\alpha = \frac{\partial \tilde {\cal L}}{\partial \dot \phi_\alpha} = \frac{\partial }{\partial \dot \phi_\alpha}\left(\tilde \pi^\beta \dot \phi_\beta - \tilde {\cal L}\right) = \tilde \pi^\alpha + \dot \phi_\beta \frac{\partial \tilde \pi^\beta}{\partial \dot \phi_\alpha} - \frac{\partial {\cal H}}{\partial \pi^\beta} \frac{\partial \tilde \pi^\beta}{\partial \dot \phi_\alpha}.
\eeq
Since $\frac{\partial \tilde \pi^\beta}{\partial \dot \phi_\alpha}$ is non-singular, we deduce that
\beq
\dot \phi^\alpha = \frac{\partial \tilde {\cal H}}{\partial \pi^\alpha}.  \label{psidot}
\eeq
Let us also calculate
\bea
\frac{\delta \tilde {\cal L}}{\delta \phi_\alpha} &=& \frac{\delta \tilde \pi^\beta}{\delta \phi_\alpha} \dot \phi_\beta - \left.\frac{\partial \tilde {\cal H}}{\partial \phi_\alpha}\right|_\pi - \frac{\delta \tilde {\cal H}}{\delta \pi^\beta}\frac{\partial \tilde \pi^\beta}{\partial \phi_\alpha}  \nonumber \\
&=&- \frac{\delta \tilde{\cal H}}{\delta \phi_\alpha},
\eea
where in the last line we used (\ref{psidot}). But according to the Euler-Lagrange equations (\ref{el}), 
\beq
\dot {\tilde \pi}^\alpha = \frac{\delta \tilde {\cal L}}{\delta \phi_\alpha},
\eeq
so we deduce that
\beq
\dot \pi^\alpha = - \frac{\delta \tilde {\cal H}}{\delta \phi_\alpha}. \label{pidot}
\eeq

We are finally prepared to state Schwinger's quantum action principle and to derive some consequences. We first assume that we have a non-singular system so that we can express the Lagrangian in terms of the operators $\phi$ and $\pi$ using (\ref{calL}). The action is then\footnote{Schwinger wrote this expression in terms of exact differentials. I will comment on the historical Hamilton-Jacobi origins in Section 8.3.}
\beq
S =\int_2^1 d^3\!x \pi^\alpha d\phi_\alpha - \int^{t_1}_{t_2}dt \int d^3\!x {\cal H}.
\eeq
The action principle states that the variations in the action operator $S$ by the independent variations $\delta \phi_\alpha$, $\delta \pi^\beta$ and $\delta x^\mu$ must depend only on variations at $t_1$ and $t_2$. The general variation is
\bea
\delta S &=& \int_2^1 d^3\!x \left(\delta \pi^\alpha d\phi_\alpha +\pi^\alpha d \delta \phi_\alpha - \int^{t_1}_{t_2} \left( \delta {\cal  H} dt + {\cal  H} d \delta t\right) \right) \nonumber \\
&=& \int^{t_1}_{t_2} dt d^3\!x \left(\delta \pi^\alpha \dot \phi_\alpha -\dot \pi^\alpha \delta \phi_\alpha -\left( \frac{\delta {\cal  H}}{\delta \phi_\alpha} \delta \phi_\alpha +\frac{\delta {\cal  H}}{\delta \pi^\beta} \delta \pi^\beta + \frac{\partial H}{\partial t} \delta t -\frac{d{\cal  H}}{dt} \delta t\right) \right) \nonumber \\
&+& \int^1_2 d \left(\int d^3\!x \pi^\alpha \delta \phi_\alpha - \int d^3\!x {\cal H} \delta t \right) \label{deltasqp}
\eea 
Thus the integrand of the time integral is required to vanish. We already have (\ref{psidot}) since this is the non-singular Legendre transformation, so we deduce the additional Hamiltonian dynamical equation (\ref{pidot}) and 
\beq
\frac{d H}{dt} = \frac{\partial H}{\partial t}.
\eeq
With these dynamical equations satisfied we obtain from (\ref{deltasqp})
\beq
\delta S_2^1 = \left.\int d^3\!x\left( \pi^\alpha \delta \phi_\alpha - {\cal H} \delta t \right)\right|_2^1.
\eeq

Let us now interpret these operators. 
We work in a representation basis of eigenstates of the Hermitian field operator $\phi(\vec x)$ with real eigenvalues labeled by primes,
\beq
\phi_\alpha(x) \left.\right| \phi' \left. \right> =\phi'_\alpha(x) \left.\right| \phi' \left. \right>.
\eeq
Then translations in the field and in the time will be generated by Hermitian operators $\int d^3\!x {\cal G}^\alpha \delta \phi_\alpha/\hbar$ and $-H\delta t/ \hbar:= -\int d^3\!x {\cal H}\delta t/ \hbar$, respectively,
\beq
\left<\right. \phi', t \left. \right| e^{i d^3x {\cal G}^\alpha \delta \phi_\alpha/\hbar} =  \left<\right. \phi' + \delta \phi', t\left. \right|,
\eeq
and
\beq
 \left<\right. \phi', t \left. \right> e^{-iH \delta t/\hbar} =   \left<\right. \phi', t + \delta t\left. \right|,
\eeq

In this representation we have therefore
\bea
&& \delta \left<\right. \phi' _1, t_1 \left. \right|  \phi'_2, t_2 \left. \right>:=\left<\right. \phi' _1+ \delta \phi'_1, t_1 + \delta t_1\left. \right|  \phi'_2+ \delta \phi'_2, t_2 + \delta t_2\left. \right> - \left<\right. \phi' _1, t_1 \left. \right|  \phi'_2, t_2 \left. \right> \nonumber \\
&=& \left<\right. \phi' _1, t_1 \left. \right|\frac{i}{\hbar}\int d^3\!x \left( {\cal G}^\alpha_1 \delta \phi_{\alpha\,1} - {\cal H}_1 \delta t_1 -{\cal G}^\alpha_2 \delta \phi_{\alpha\,2} +{\cal H}_2 \delta t_2 \right)\left. \right|  \phi'_2, t_2 \left. \right>\nonumber \\
&=&   \left<\right. \phi' _1, t_1 \left. \right| \frac{i}{\hbar}\left( \left.\int d^3\!x\left( \pi^\alpha \delta \phi_\alpha - {\cal H} \delta t \right)\right|^2_2 \right) \phi'_2, t_2 \left. \right>
\eea

We deduce that ${\cal G}^\alpha(x) = \pi^\alpha(x)$. But since
\beq
e^{i d^3\!y {\cal G}^\alpha(y) \delta \phi'(y)_\alpha/\hbar} \phi_\beta(x) e^{-i d^3\!z {\cal G}^\alpha(z) \delta \phi'(z)_\alpha/\hbar} = \phi_\beta(x) + \delta \phi'_\beta(x) = -\frac{i}{\hbar} \left[\phi_\alpha(x),  d^3\!y {\cal G}^\beta(y)\right],
\eeq
we know that
\beq
\left[\phi_\alpha(x),  {\cal G}^\beta(y)\right] =i \hbar \delta^\beta_\alpha \delta^3 (x-y).
\eeq
We deduce therefore  that $\phi$ and $\pi$ satisfy the commutation relations
\beq
\left[\phi_\alpha(x),  \pi^\beta(y)\right] =i \hbar \delta^\beta_\alpha \delta^3 (x-y).
\eeq
Note furthermore, as a consequence of these commutation relations,
\beq
\left[\phi_\alpha(x), F[\phi,\pi]  \right] = i \hbar \frac{\delta F}{\delta \pi^\alpha(x)},
\eeq
and
\beq
\left[\pi^\alpha(x), F[\phi,\pi]  \right] = -i \hbar \frac{\delta F}{\delta \phi_\alpha(x)}.
\eeq

Now, regarding the time development of operators in the Heisenberg representation, we have
\beq
\phi_\alpha(x,t-\delta t) = e^{-i H \delta t/\hbar} \phi_\alpha(x,t) e^{i H \delta t/\hbar} = \frac{i}{\hbar} \left[ \phi_\alpha(x,t), H \right] = \phi_\alpha(x,t) - \frac{\delta H}{\delta \pi^\alpha(x)} \delta t,
\eeq
so we recover the equation of motion (\ref{psidot}). In the same manner we also recover the equations of motion (\ref{pidot}). We have the remarkable result that the quantum equations of motion can be simply read off the the Hamilton-Jacobi action variation $\delta S^2_1$. 

It is noteworthy that given an expression for the generator of variations of the action of the form 
\beq
G_q = \frac{1}{2}\left(p_k  \delta q^k - \delta q^k p_k  \right) - H(q,p,t) \delta t,
\eeq
the $q$ and $p$ satisfy canonical commutation relations, and they satisfy canonical Hamiltonian equations of motion.\footnote{See \cite{Schwinger:1953ab} for a compact discussion of the more general case in which anti-commutation relations arise.} Obviously, one can also obtain from this expression a corresponding Lagrangian, just by inverting process that we described above for a non-singular system.

\subsection{Schwinger's quantum electrodynamics}

In the preceding section we assumed that the field theory was non-singular. However, Schwinger did discuss in his 1951 article how to deal with a special class of singular theories, namely those for which the Lagrangian density did not depend on a subset of velocities. He then represented the subset of configuration variables for which the definitions of canonical momenta could be inverted to give the velocities as functions of momenta by $\phi_a$. Those variables whose velocities did not appear in the Lagrangian density were represented by $\phi_A$. He did insist that all operator matrix elements be invariant under the resulting gauge symmetry group. In particular, he insisted that the operator eigenstates be gauge invariant.

His procedure for dealing with electrodynamics was in fact to solve the constraints and gauge conditions, where he effectively employed the Coulomb gauge. The innovation in his operator approach was to begin with a Lagrangian density that was linear in velocities. In this form one can simply ``read off'' the variation of the action generator. We refer to the second of his series of articles on the quantum action principle \cite{Schwinger:1953aa}.\footnote{See also \cite{Schwinger:2001aa}  and for the classical approach \cite{Schwinger:1998aa}.}

Schwinger takes as his free electromagnetic Lagrangian density (where $\left\{ F,G \right\} := FG + GF$)
\bea
{\cal L}_{S\,em} &=&\frac{1}{4} \left\{F^{\mu \nu}, \partial_\mu A_\nu -\partial_\nu A_\mu \right\}  -\frac{1}{4} F^{\mu \nu} F_{\mu \nu} \nonumber \\
&=& \frac{1}{2} \left\{F^{0a}, \frac{1}{c} \dot A_a - A_{0,a} \right\} + \frac{1}{2} \left\{F^{ab}, A_{b,a}\right\} -\frac{1}{2} F^{0a} F_{0a} -\frac{1}{4} F^{ab} F_{ab}.
\eea
It is assumed that $F^{(\mu \nu)} = 0$ and it is independent of $A_\mu$. We then have
\beq
\tilde \pi^a := \frac{\partial {\cal L}_{S\,em}}{\partial \dot A_a} = F^{0a},
\eeq
and the Euler-Lagrange equations are
\beq
\dot F^{0a} =- \partial_b F_{ba} = - \left( \vec \nabla \times \left(\vec \nabla \times \vec A\right)\right)_a ,
\eeq
\beq
F^{0a} = \frac{1}{c}\left( \dot A_a + V_{,a} \right),
\eeq
\beq
F_{ab} = \partial_a A_b - \partial_b A_a, \label{fab}
\eeq
and
\beq
F^{0a}_{,a} = 0.
\eeq

The variation of the action at the limits of the time integration is
\beq
\delta S_{S\,em} = \int d^3\!x \left(\pi^a \delta A_a - {\cal H}_{S\,em} \delta t -  {\cal P}_{S\,em\,a} \delta x^a  \right),
\eeq
where
\beq
{\cal H}_{S\,em} = \pi^a \dot A_a - {\cal L}_{S\,em}, \label{hsem}
\eeq
and
\beq
{\cal P}_{S\,em\,a} = \pi^b A_{b,a}. 
\eeq
We note that even though the definition of the momentum conjugate to $A_a$ cannot be solved for $\dot A_a$ in terms of $\pi^b$, this is of no consequence since the $\dot A_a$ terms cancel in (\ref{hsem}), and we find that
\beq
{\cal H}_{S\,em} =  -\pi^a A_{0,a} + \frac{1}{2} \left\{F^{ab}, A_{b,a}\right\} +\frac{1}{2} \pi^a \pi_a -\frac{1}{4} F^{ab} F_{ab}
\eeq

It is at this stage that Schwinger solves the constraints and gauge conditions so as to obtain a gauge invariant variation of the action operator. In addition to the secondary constraint $F^{0a}_{,a} = 0$ from the conventional analysis, there is an additional constraint resulting from the use of the $F_{ab}$ auxiliary variable. The procedure as he describes it is to simply eliminate those degrees of freedom for which there are no dynamical equations. The result is the equivalent of the gauge fixing procedure that we described earlier. He eliminates the auxiliary field $F^{ab}$ using (\ref{fab}), imposes the Coulomb gauge condition $\vec \nabla \cdot \vec A = 0$, thereby also eliminating $A_0$.

\section{The reduced phase space of the Bergmann school and temporal evolution} \label{bergmann}

I turn now to the spacelike hypersurface integrals that arise in the variation of the gravitational field action. Peter Bergmann and his collaborators examined in detail both general canonical variations and those variations that arise through general coordinate transformations \cite{Bergmann:1949aa,Bergmann:1949ab,Bergmann:1949ac,Bergmann:1950ab,Anderson:1951aa}, and I will only briefly summarize their results.\footnote{Some had already been obtained by Rosenfeld \cite{Rosenfeld:1930aa} \cite{Rosenfeld:2017ab}, as discussed in \cite{Salisbury:2017aa}. See also \cite{Blum:2018aa} for an analysis of this earlier work.} One specific pertinent fact here is that when attention is confined to variations under coordinate transformations the resulting phase space generators vanish - excepting for possible two-surface integrals at spatial infinity. The vanishing generators were in fact linear combinations of phase space constraints. The outcome was that these generators could be employed to generate distinct solutions of Einstein's field equations from identical initial conditions. They argued therefore that all solutions obtained in this manner corresponded to the same physical state. There was no way of physically distinguishing them. According to them true physical observables needed to be invariant under those variations generated by the constraints. Indeed, their definition of a gravitational observable was that it was a phase space functional that was invariant under the action of the constraints. 

First I address variations of the action under an arbitrary infinitesimal general coordinate transformations $x'^\mu = x^\mu - \epsilon^\mu(x)$. Bergmann denotes generic fields by $y_A$ and he represents the variation under the coordinate transformation as $y'_A(x) - y_A(x)=: \bar \delta y_A(x)$. This is actually the Lie derivative with respect to $\epsilon^\mu(x)$.  He assumes that the Lagrangian density $ \mf{L}$ plus an additional total divergence term $S^\mu_{,\mu}$ transforms as a scalar density of weight one,
\beq
\bar \delta \mf{L} \equiv- \left(\bar \delta S^\mu \right)_{, \mu} + \left(\left(\mf{L}+ S^\nu_{,\nu}\right)\epsilon^\mu\right)_{,\mu} =  \mf{L}^A \bar \delta y_A + \left(\frac{\partial  \mf{L}}{\partial y_{A, \mu}} \bar \delta y_A \right)_{, \mu}, \label{ident}
\eeq
where $\mf{L}^A = 0 $ are the Euler-Lagrange equations. Furthermore, he and his collaborators (as did Rosenfeld) assumed that  $\mf{L}$ did not depend on $y_{A,\mu \nu}$ or higher derivatives, and that $S^\mu  = f^{A \mu \rho}(y) y_{A,\rho}$. It follows from (\ref{ident}) that when the field equations are satisfied, defining 
\beq
\mf{C}^\mu := -\bar \delta S^\mu - \left(\mf{L}+ S^\nu_{,\nu}\right)\epsilon^\mu - \frac{\partial  \mf{L}}{\partial y_{A, \mu}} \bar \delta y_A, \label{cmu}
\eeq
that $\mf{C}^\mu_{, \mu} = 0$. Choosing vanishing variations at spatial infinity we conclude that 
therefore  $\frac{d}{dt} \int d^3\!x\,\mf{C}^0 = 0$. Since the corresponding temporally constant spatial integral, $\int d^3x \,\mf{C}^0$ contains the arbitrary descriptors $\epsilon^\mu$ the the coefficients of each time derivative must itself vanish.\footnote{This argument was actually first formulated by Rosenfeld \cite{Rosenfeld:1930aa,Rosenfeld:2017ab}. See \cite{Salisbury:2017aa} for a detailed analysis.} This is indeed the vanishing Noether charge whose existence follows from Noether's second theorem. Anderson and Bergmann showed that the first time derivative was the highest that appeared, i.e. $\mf{C}^0 = {}^0A_\mu \epsilon^\mu + {}^1A_\mu \dot \epsilon^\mu$, and that the functions $ {}^1A_\mu $ were the primary constraints that arose from the momentum definition $\pi^A := \frac{\partial \mf{L}}{\partial \dot y_A}$. The absence of higher time derivatives was a puzzle, though not mentioned explicitly in writing by Bergmann until much later. There existed an apparent conflict with nested diffeomorphism Lie algebras since these would lead to higher time derivatives\footnote{see \cite{Salisbury:2020aa} for further details.}. This puzzle was partially resolved, though perhaps inadvertently, by Paul Dirac in his Hamiltonian formulation of general relativity \cite{Dirac:1958aa}. Bergmann \cite{Bergmann:1962ac} interpreted Dirac's achievement as a statement that infinitesimal general coordinate transformation in the Hamiltonian formalism were restricted to depend on the lapse $N$ and shift $N^a$ components of the metric
\beq
g_{\mu \nu} = \left(\begin{matrix}N^2 - g_{ab} N^a N^b & g_{ab}N^b\\g_{ab}N^b & g_{ab}\end{matrix} \right).
\eeq
Specifically, the permissible infinitesimal transformations involved descriptors of the form 
\beq
\delta x^\mu =- \epsilon^\mu(x, N^\mu)= -n^\mu \xi^0(x) - \delta^\mu_a \xi^a(x), \label{deltax}
\eeq
 where $n^\mu = \left(N^{-1}, -N^{-1} N^a \right)$ is the normal to the constant time hypersurface. In addition, Bergmann showed that the infinitesimal commutator algebra obeyed by these transformations was of the form
\beq
\xi^\rho = \delta^\rho_a \left(\xi^a_{1,b} \xi^b_2 - \xi^a_{2,b} \xi^b_1 \right) + e^{ab} \left(\xi^0_1 \xi^0_{2,b} - \xi^0_2 \xi^0_{1,b}\right) + n^\rho \left(\xi^a_2 \xi^0_{1,a} - \xi^a_1 \xi^0_{2,a}\right). \label{dalgebra}
\eeq
In Dirac's formalism it was clear to Bergmann (although Dirac never made this claim) that the secondary constraints
\beq
{\cal H}_0 =  \frac{1}{\sqrt{g}} \left(p_{ab}p^{ab} - (p^c{}_c)^2) \sqrt{g} {}^3\!R \right) = 0, \label{h0}
\eeq
 and 
 \beq
 {\cal H}_a = - 2 p^b{}_{a|b} = 0, \label{ha}
 \eeq
through the spatial integrals $\int d^3x \xi^\mu  {\cal H}_\mu $, generated the correct variations of $g_{ab}$ and $p^{cd}$ corresponding to the descriptors $\epsilon^\mu$ in (\ref{deltax}). (In these expressions $g$ is the determinant of the 3-metric, $p^{ab}$ the momentum conjugate to $g_{ab}$, indices are raised or lowered with the 3-metric, and the $|$ signifies covariant derivative in the 3-space.) Curiously, Bergmann did not give the Poisson bracket algebra that corresponds to the descriptor algebra in (\ref{dalgebra}). This algebra subsequently became known as the Dirac algebra, although as far as I can tell Dirac never published a computation of it employing the secondary constraints $ {\cal H}_\mu$ of general relativity. The designation comes from his flat spacetime example in which he worked with curved spatial hypersurfaces \cite{Dirac:1964aa}.  Higgs did give a portion of it \cite{Higgs:1958aa,Higgs:1959aa}, but he omitted the crucial bracket that contained the metric field dependence,
$$
\left\{{\cal H}(x)_0 , {\cal H}(x')_0 \right\} =  \int d^3 x'' {\cal H}_a(x'') e^{ab}(x'') \left(\delta^3(x-x'') + \delta^3(x'-x'')\right) \frac{\partial}{\partial x^b} \delta^3 (x-x')
$$
\beq
=:  {\cal H}_{a''} C^{a''}_{0 0'}, \label{hlhl}
\eeq
where I introduce the convention that repeated primed indices also imply a spatial integration. I represent the full algebra as
\beq
\left\{{\cal H}_{\mu'}, {\cal H}_{\rho''} \right\}=: C^{\nu'}_{\mu' \rho''} {\cal H}_{\nu'}.
\eeq

It is remarkable that as a consequence of Dirac's work Bergmann abandoned the inclusion of the lapse and shift as canonical phase space variables.\footnote{Joshua Goldberg has informed me that Dirac was the direct cause.} Thus in his subsequent work the $\dot \xi$ terms were missing, and as a consequence Bergmann was apparently not concerned with demonstrating the full four-dimensional covariance of the Hamiltonian formulation of general relativity, even though he assumed its appearance in the notion to be introduced shortly of a physical reduced phase space. 

Bergmann and his collaborators also considered variations of solutions of Einstein's that did not necessarily correspond to changes in coordinates. Bergmann and Schiller \cite{Bergmann:1953aa} focused on those variations in configuration-velocity space that would not change the form of the field equations, and could be expressed in terms of phase space variables. They observed that generally these variations involved an additional total divergence in the Lagrangian. Bergmann and I. Goldberg continued this investigation, and in the process invented a new Poisson bracket. They called it `an extend Dirac bracket', but its group theoretical origins and unique functional form deserves recognition, and I have assigned it the name `Bergmann-Goldberg' bracket \cite{Salisbury:2020aa}. It expressly represents the algebra of observables that are invariant under the action of the full 4-dimensional diffeomorphism group. This is the algebra of the reduced phase space which is obtained by quotienting the general transformations that retain solution fields by the diffeomorphism invariance group. It does appear at first sight that this algebra is unique in that it does not appear to involve the imposition of coordinate (gauge) conditions. But the scheme does require a choice of null vectors of the symplectic. Such a selection is indeed equivalent to Dirac brackets when the gauge freedom is fully eliminated.\footnote{See \cite{Salisbury:2020aa}}

The resulting observable fields which satisfy the reduced algebra are diffeomorphism invariants and therefore constants of the motion. The Syracuse school explored two related means to obtain them. One involved the imposition of coordinate conditions, and I have explored the earlier history of this approach in detail in \cite{Salisbury:2020aa}. Bergmann and Komar were also apparently the first to propose a canonical change of variables such that invariants could undergo non-trivial evolution in an intrinsic time. They spelled this out in some detail \cite{Bergmann:1962ad} in a contribution to the 1962 Infeld Festschrift that was evidently composed prior to April, 1960.\footnote{The article contains a note added in proof referring to a work  \cite{Bergmann:1960aa} that they submitted for publication on March 31, 1960.} The idea they presented was to isolate an invariant variable $C$ in the Hamiltonian constraint $H=0$, perhaps through a canonical change of variable. The canonical conjugate $\theta$ to $C$ could not appear in this constraint since by its definition as an invariant the Poisson bracket of $C$ with $H$ vanishes. This $\theta$ then assumes the role of evolution time. It will be shown in detail in \cite{Salisbury:2021ad}  how this occurs for the free relativistic particle.

\section{The ADM formalism and Kucha\v{r}} \label{ADM}

Richard Arnowitt and Stanley Deser were Schwinger's doctoral students at Harvard in the early 1950's. Their first foray into quantum gravity in 1959 \cite{Arnowitt:1959ab} was directly inspired by Schwinger's quantum action principle. As had Schwinger in the electromagnetic case, they argued that quantum transition amplitudes could be sensibly defined only between states characterized by gauge invariant physical quantities. Applied to quantum gravity they recognized that the Schwinger action principle necessarily dealt with invariants under the diffeomorphism symmetry group. Already in their first paper they identified as a useful analogue of Schwinger's first-order analysis of electromagnetism an approach to general relativity in which the metric and the Christoffel connection were taken to be independent fields. After their initial work in which they worked out the implications of their approach in linearized gravity, they joined forces with Charles Misner. There followed a series of papers by Arnowitt, Deser, and Misner (ADM) that culminated in a comprehensive overview in 1962 \cite{Arnowitt:1962aa}. I will refer mainly to this overview in this discussion. 

Their first-order Palatini approach afforded them a framework where they could quickly identify an appropriate alteration of the Hilbert action of general relativity that would simplify the primary constraints of the theory. They were aided in this search through their use of lapse and shift variables and their identification of the extrinsic curvature as dynamical phase space variables. This choice was itself a natural consequence of the first order formalism. Their procedure led to a Lagrangian that differed by a divergence both from Dirac's, and the original first order form considered by Einstein and Bergmann. It took the form 
\beq
\mf{L}_{ADM} = N \sqrt{g}\left({}^3\!R + K_{ab}K^{ab} -\left(K^a{}_a\right)^2 \right), \label{ADM}
\eeq
where $g$ is the determinant of the spatial metric. In this form the extrinsic curvature is conceived as a function of the lapse and shift, and takes the form $K_{ab} = \frac{1}{2N} \left(\dot g_{ab} - N_{a|b} - N_{b|a}\right)$. The full variation of this Lagrangian as a function of the metric and its derivatives, and the resulting constraints will be dealt with in \cite{Salisbury:2021ad}, but at the moment I will concentrate on the non-vanishing temporal boundary terms that were obtained by ADM, also in their first order formalism,
 namely
\beq
\delta S = \int d^3\!x \, p^{ab}\delta g_{ab}, \label{deltasadm}
\eeq
where $p^{ab} = \sqrt{g} \left( K^{ab} - K^c{}_c e^{ab}\right)$, with indices raised by the inverse spatial metric $e^{ab}$.
This arises as an example of field variations considered by Weiss, and discussed in Section 3.  We shall continue to interpret these variations as representing variations to new solutions at the fixed time $t$. ADM proposed to introduce gauge conditions in this expression, and this is precisely the program that is pursued in \cite{Salisbury:2021ad}  ADM also correctly recognized that the introduction of gauge conditions with explicit reference to the coordinates is equivalent to a choice of ``intrinsic coordinates"\footnote{\cite{Arnowitt:1962aa}, p. 232}, and that the intrinsic coordinates must be scalar spacetime functions of the variables $g_{ab}$ and $p^{cd}$. (The advance discussed in \cite{Salisbury:2021ad} is that the corresponding spacetime diffeomorphism invariants can be explicitly displayed.)
To be more specific, ADM proposed to introduce gauge conditions $x^\mu = q^\mu[g_{ab}, p^{cd}]$, where the $q^\mu$ are spacetime scalars.  In this manner the metric itself individuates locations in spacetime.  Indeed, already in the second of their jointly authored papers \cite{Arnowitt:1960aa} they suggested as a possible choice the spacetime scalars that arise when one imposes de Donder (harmonic) gauge  conditions $\left[\left( -g \right)^{1/2} g^{\mu \nu}\right]_{,\nu} = 0$. These four scalars $\bar v^\rho$ are the four linearly independent solutions of 
\beq
\bar v^\mu_{;\nu}{}^{;\nu} := \left( - {}^4 g\right)^{-\frac{1}{2}} \left[\left( - {}^4 g\right)^{\frac{1}{2}} g^{\mu \nu }\bar v^\rho_{,\nu}\right]_{, \mu}=0.
\eeq
These scalars will be non-local functionals of the metric. It is remarkable that ADM did not here refer to Komar and Bergmann who had earlier proposed the use of Weyl curvature scalars in the construction of intrinsic coordinates.\footnote{Indeed, in remarks made in response to Bergmann's talk at the Royaumont meeting in June, 1959 \cite{Bergmann:1962aa}, they offered objections to the use of Weyl scalars with reasoning that would seem to apply also to their own coordinate conditions.} In fact, they are the only local scalars that can be built with the vacuum Riemann curvature. Bergmann and Komar had shown in 1960 that these scalars depended only on $g_{ab}$ and $p^{cd}$ \cite{Bergmann:1960aa}. 

\subsection{Kucha\v{r} approach}

Karel Kucha\v{r} proposed initially in 1972 \cite{Kuchar:1972aa} that the physical degrees of freedom of vacuum gravity could be isolated through an appropriate canonical transformation of the variables $g_{ab}$ and $p^{cd}$. He was likely not aware that Bergmann and Komar had proposed a similar procedure over ten years earlier as I noted at the end of Section 4. Kucha\v{r} continued with a series of papers in which the implications of this approach were examined \cite{Kuchar:1976aa,Kuchar:1976ab,Kuchar:1976ac,Kuchar:1981ab}. I will employ the notation used by Isham in his overview of the Kuchar program \cite{Isham:1993aa}.

Kucha\v{r} was the first to contemplate a canonical transformation to new variables $X^\mu$ that could be interpreted as ``internal" space and time coordinates, 
\beq
\left(g_{ab}(x), p^{cd}(x)  \right) \rightarrow \left( X^\mu(g_{ab}(x), p^{cd}(x) ), {\cal P}_\nu(g_{ab}(x), p^{cd}(x) );\phi^r(x),\pi_s(x)  \right).
\eeq
The $ {\cal P}_B(x)$ are their conjugate momenta. The remaining two canonical pairs $\phi^r(x),\pi_r(x) $, with $r=1,2$ are to represent the true physical degrees of freedom of the gravitational field. Kucha\v{r} then supposed that the gravitational action took the form
\beq
S_K[\phi,\pi,N^\mu,X,{\cal P}] := \int dt \int d^3\!x \,\left({\cal P}_A \dot X^A + \pi_r \dot \phi^r - N^\mu {\cal H}_\mu  \right). \label{sk1}
\eeq
One then solves the constraints for ${\cal P}_A $ in the form
\beq
{\cal P}_A = - h_A \left[x; X, \phi, \pi  \right].
\eeq
Furthermore, the functions $X$ are set equal to prescribed functions of the original canonical variables,
\beq
X^A(g,p) = \chi^A(x).
\eeq
(To be consistent this solution cannot depend on the variables ${\cal P}$ that have already been eliminated.) 
One thereby obtains
\beq
{\cal P}_A(x, \chi; \phi, \pi) = - h_A \left[( \chi; \phi, \pi)  \right].
\eeq 
Then one substitutes the solutions into (\ref{sk1}) yielding
\beq
S_K[\phi,\pi]=  \int dt \int d^3\!x \left\{ \pi_r(x) \dot \phi^r (x)- h_A\left[( \chi; \phi, \pi)  \right] \dot \chi^A(x)\right\}. \label{sk2}
\eeq
In particular one could choose as intrinsic coordinates $\chi^\mu(x) = x^\mu$, in which case the action takes the form
\beq
S_K[\phi,\pi]=  \int dt \int d^3\!x \left\{ \pi_r(x) \dot \phi^r (x)- h_0\left[( x; \phi, \pi)  \right] \right\}. \label{sk3}
\eeq
The canonical Hamiltonian equations of motion for $\phi$ and $\pi$ then follow, with the Hamiltonian density $h_0\left[( x; \phi, \pi)  \right] $. A variant of this approach will be pursued in \cite{Salisbury:2021ad} .

\section{Bergmann and Komar's Hamilton-Jacobi approach} \label{BKHJ}
 
It is a first sight puzzling that Bergmann, and subsequently also Komar, turned their attention to the classical Hamiltonian-Jacobi treatment of general relativity. But as mentioned above, we have the evidence from his two textbooks \cite{Bergmann:1949ac,Bergmann:1951ab} that Bergmann was certainly cognizant of the formalism and its link to quantum theory.  I note also that Freistadt in 1955 in his paper on the Hamilton-Jacobi treatment of classical field theory thanks Bergmann for ``stimulating discussions".\footnote{\cite{Freistadt:1955aa}, p. 1161.} Perhaps less appreciated is Peres' expression of gratitude to Bergmann, in his groundbreaking 1962 paper, for Bergmann's observation that Peres' functional $S$  could be interpreted ``as the Hamilton-Jacobi functional for the gravitational field."\footnote{\cite{Peres:1962aa}, p. 60.}  In Bergmann's initial Hamilton-Jacobi paper in 1966 \cite{Bergmann:1966aa} he  did make a reference to his own suggested Hamilton-Jacobi inspired Schroedinger approach in his short abstract for a 1950 meeting \cite{Bergmann:1952aa}. This is followed by a citation of a full-scale Hamilton-Jacobi quantization proposed by Shanmugadhasan \cite{Shanmugadhasan:1963aa}.  It is curious that this short paper makes no reference to Wheeler, nor to Peres. In fact, as  we shall see, this study is clearly motivated by the previous work of Bergmann and Komar in attempting to determine the structure of the reduced phase space of classical general relativity. This project was simply not at the forefront of the competing schools. 

The Hamilton-Jacobi approach to general relativity was of course severely challenged by the underlying general covariance of the theory.  As we shall see explicitly in \cite{Salisbury:2021ad}, the classical Hamilton-Jacobi variations, modeled after Weiss and Schwinger, yield contributions that are constrained to vanish. How is one to take them into account? The proposal advanced by Peres \cite{Peres:1962aa} was to assume that a principle characteristic functional $S[g_{ab}, x]$ could be found such that the three-momentum satisfied $p^{ab} = \frac{\delta S}{\delta g_{ab}}$ - and this functional should satisfy the four secondary constraints (\ref{h0}) and (\ref{ha}),
\beq
{\cal H}_\mu \left( g_{ab},\frac{\delta S}{\delta g_{cd}} \right) = 0. \label{hjeqs}
\eeq
 Although these equations have collectively become known as `Hamitlon-Jacobi' equations, this is a stretch in terminology. True, if $S$ may be interpreted as the action, then according to (\ref{deltasadm}) the three momentum may be expressed as a functional derivative. But could a complete  solution for $S$ still be interpreted as generating a canonical transformation to a temporally non-evolving phase space? Could general solutions of Einstein's equations be obtained by performing appropriate functional derivatives, and perhaps most importantly, could this feature be employed in a Schr\"odinger framework to produce solutions of Einstein's equations in a semi-classical regime? 

Bergmann obtained in his first paper a result that was consistent with his earlier canonical approach. He proved that $S$ satisfying a finite dimensional analogue of (\ref{hjeqs}), with  first class constraints $C_a(q,p)=0$ and Hamiltonian $H = \eta^a C_a$, could contain no explicit dependence on the time coordinate $t$, and this was a reassuring indication that $S$ in general relativity might involve the diffeomorphism invariants that featured in the phase space approach. This lent credence to the expectation that the reduced phase space algebra could be obtained in this formalism. This is the clear rationale for taking up the Hamilton-Jacobi analysis. Indeed, Bergmann noted in this regard that ``\ldots the Hamilton-Jacobi theory does not require the setting of `gauge conditions,' such as is implicit in any choice of the coefficients $\eta^a$ in the expression for the Hamiltonian \ldots", and we shall see how this comes about. The conclusion that there is no coordinate time dependence comes from a simple argument referring to the finite dimensional system. It is easily generalized to exclude spatial coordinate dependence as Komar noted in  \cite{Komar:1967aa} . Bergmann has
$$
\dot p_k = -\frac{\partial H}{\partial q^k} = \frac{d}{dt} \frac{\partial S}{\partial q^k} = \frac{\partial^2 S}{\partial q^k \partial q^l} \dot q^l + \frac{\partial^2 S}{\partial q^k \partial t} = \frac{\partial^2 S}{\partial q^k \partial q^l} \frac{\partial H}{\partial p_l}+ \frac{\partial^2 S}{\partial q^k \partial t},
$$
or 
\beq
\frac{\partial H}{\partial q^k}  +  \frac{\partial^2 S}{\partial q^k \partial q^l} \frac{\partial H}{\partial p_l}+ \frac{\partial^2 S}{\partial q^k \partial t} = 0. \label{pdot}
\eeq
But the $p$ that appears in $H$ must be understood as a function of $q$, and as such a function it must vanish identically since it is a linear combination of the constraints, i.e., $H = \eta^a C_a\left( q, \frac{\partial S(q,t)}{\partial q} \right) \equiv 0$ since $S$ explicitly solves the constraint equations. In other words
\beq
0 = \frac{\partial H(q, p(q))}{\partial q^k} = \frac{\partial H}{\partial q^k} + \frac{\partial^2 S}{\partial q^k \partial q^l} \frac{\partial H}{\partial p_l}. \label{Hofq}
\eeq
Therefore, according to (\ref{pdot}) and (\ref{Hofq}), $\frac{\partial}{ \partial q^k}\frac{\partial S}{ \partial t} = 0$, or $\frac{\partial S}{ \partial t} = f(t)$. But this "constant" of integration will have not effect on the dynamics, so can as well take $\frac{\partial S}{ \partial t} = 0$. Although Bergmann did not write this, the analogue of the same argument is valid in general relativity, where in principle $S$ could depend explicitly on the spatial coordinates $\vec x$, i.e., one might have $S[g_{ab}(\vec x);\vec x]$. Komar noted in \cite{Komar:1967aa}  that indeed, one had $S[g_{ab}(\vec x)]$ and he attributed this result to Bergmann. It is also important to note that $g_{ab}(\vec x)$ has no time dependence. Also, although Bergmann did not give the details\footnote{They were later provided by Komar \cite{Komar:1967aa}.}, he did recognize that the homogeneity properties of the Hamiltonian constraints could be exploited to deduce the variation of $S$ under diffeomorphisms. In particular, given that the variation in $g_{ab}$ under an infinitesimal coordinate change is given by $\delta g_{ab} = \frac{\delta}{\delta p^{ab}} \left(\epsilon^\mu {\cal H}_\mu \right) $, since the local change in $S$ is given by 
$\delta S = \frac{\delta S}{\delta g_{ab}} \delta g_{ab}= p^{ab} \frac{\delta}{\delta p^{ab}} \left(\epsilon^\mu {\cal H}_\mu \right) $, the homogeneity in $p^{ab}$ can be employed to find the change in $S$.\footnote{As Komar observed, one must assume either the the spatial manifold is closed, or appropriate asymptotic conditions.} Then since the constraints ${\cal H}_a$ are homogeneous of degree one, the change is proportional to the constraint, and therefore vanishes. Furthermore, the terms with homogeneity degree two in ${\cal H}_0$ may be exploited to find the change in $S$ under infinitesimal temporal displacements, resulting in 
$\delta S = -2 \epsilon^0  \left({}^3 g\right)^{1/2}{}^3 R$.\footnote{The square root was missing in Bergmann's account, and was added by Komar in 1967 \cite{Komar:1967aa}. } Thus it turned out that $S$ is not necessarily constant within an equivalence class. Komar later addressed the implications of this result.

Next, in a first step in deducing the invariant algebra, Bergmann considered an infinitesimal canonical transformations of $S$ with generator $A(q,p)$, so that $\delta q^k = \frac{\partial A}{\partial p_k}$ and $\delta p_k = -\frac{\partial A}{\partial q^k}$. He then asked what would be the nature of generators that did not alter the $p_k(q) = \frac{\partial S}{\partial q^k}$.
The functional form of the change of $p$ as a function of its argument $q$ is
\beq
\delta' p_k = \delta p_k - \frac{\partial p_k}{\partial q^l} \delta q^l =
-\left(\frac{\partial A}{\partial q^l} + \frac{\partial^2 S}{\partial q^l \partial q^k}\frac{\partial A}{\partial p_l} . \right) 
\eeq
Thus the change in the functional form of $p^k$ is minus the gradient of $A\left(q,\left(\frac{\partial S}{\partial q}\right)\right)$. It follows that the change in the functional form of $S$ is 
\beq
\delta'S = - A\left(q,\frac{\partial S}{\partial q}\right), \label{deltaprime}
\eeq
and therefore the form of the corresponding Hamilton-Jacobi equations would be unchanged when the gradient of $A(q,p(q))$ is non-zero. In other words, these differential equations retain their form under canonical transformations generated by constants of the motion. The vanishing constraints $C_a$ therefore preserve this form - but in addition so do phase space functions that are invariant under time evolution. Thus the constraints generate transformations within an equivalence class while diffeomorphism invariants generate changes between equivalence classes. This is displayed in Appendix B for the free relativistic particle.

Bergmann concluded this article with a comment on the `formal Schr\"odinger theory'. Since the Hamiltonian $H$  is a linear combination of constraints, it would follow from $H \Psi = i \hbar \frac{\partial \Psi}{\partial t} = 0$ that ``the Heisenberg and Schr\o dinger pictures are indistinguishable in any theory whose Hamiltonian is a constraint." But perhaps the preferable course, which avoids nonphysical states and factor ordering problems, is to work with classical invariants. But then, in a final remark,  ``there arises the problem of constructing useful observables."

Komar began has 1967 \cite{Komar:1967aa} article with further analysis of $S$ in general relativity, following up on Bergmann's results.  Komar went on to argue that the previously cited temporal change within an equivalence class was illusory and was not in conflict with the proven frozen nature of time. He highlighted one aspect of the absence of temporal development in 1971 \cite{Komar:1971aa} where he showed that a canonical transformation could be undertaken that rendered all four of the constraints homogeneous in the transformed momentum. I return below to the critical implications of this result.

The primary objective of the 1967 paper was to identify complete solutions of the Hamilton-Jacobi equations for general relativity. Following up on Bergmann's observation, Komar found that these solutions are indeed characterized by a commuting set of $2 \times \infty^3$ constants of the motion (diffeomorphism invariants) which in a subsequent paper \cite{Komar:1968aa} he called $\alpha^0_1[g_{ab},p^{cd}]$ and $\alpha^0_2[g_{ab},p^{cd}]$, with the notation suggesting the parallel with conventional Hamilton-Jacobi theory. He argued that although $S[g_{ab}(\vec x)]$ delivers $6\times \infty^3$ $ p^{ab}(\vec x)$'s, $S$ satisfies only $4\times \infty^3$ Hamiltonian-Jacobi equations. Therefore the $p^{ab}$ are not fully determined. He therefore added $2\times \infty^3$ $ p^{ab}(\vec x)$ additional Hamilton-Jacobi equations 
\beq
H_{A'}\left[g_{ab}(\vec x) ,  \frac{\delta S}{\delta g_{cd}(\vec x)}\right] =0, \label{alphaa}
\eeq
where $A' = 1,2$. (The $\vec x$ appearing in these expressions is to be interpreted as signifying a continuous field index, and in the interest of compactness will be omitted in some of the following expressions.) He denoted the resulting six constraints by $H_A$. One then concludes that all six constraints must have vanishing Poisson brackets with each other since as a consequence of
$$
\frac{\delta H_A}{\delta g_{ab}} +  \frac{\delta H_A}{\delta p^{cd}}  \frac{\delta^2 S}{\delta g_{cd} \delta g_{ab} } = 0,
$$
we find
$$
\left\{H_A, H_B \right\} = \frac{\delta^2 S}{\delta g_{ab} \delta g_{cd}} \left(- \frac{\delta H_A}{\delta p^{ab}}  \frac{\delta H_B}{\delta p^{cd} }+ \frac{\delta H_A}{\delta p^{cd}}  \frac{\delta H_B}{\delta p^{ab} }\right) = 0.
$$
Thus the additional phase space variables must be invariant under transformations generated by the ${\cal H}_\mu$. Komar represented the numerical values by $\alpha_A(\vec  x)$, i.e.,
\beq
\alpha^0_A\left[g_{ab}( \vec x) ,  \frac{\delta S}{\delta g_{cd}( \vec x)}\right] =\alpha_A( \vec x), \label{alphaa}
\eeq
 for $A=1,2$. Furthermore, the presumed independence of the $\alpha^0_A$ implies that 
$\left[ \alpha^0_A, g_{ab} \right] \neq 0$. ``Thus, a family of solutions, as well as its associated Hamilton-Jacobi functional $S[g_{ab}]$, determines, and is uniquely determined by, a complete commuting set of constants of the motion (observables). All the Riemannian-Einstein manifolds of a family have $2\times\infty^3$ observables differing from manifold to manifold." In this article Komar introduced the vanishing invariants\footnote{I have changed the symbol in an attempt to avoid some confusion that could result from Komar's notation. Komar called this vanishing invariant $\alpha_A\left(\vec x; g_{ab}( \vec x) ,  \frac{\delta S}{\delta g_{cd}}( \vec x) \right) $, \cite{Komar:1968aa}, equation (1.8).}
\beq
\bar \alpha_A\left(\vec x; g_{ab}( \vec x) ,  \frac{\delta S}{\delta g_{cd}}( \vec x) \right) := \alpha^0_A\left[g_{ab}( \vec x) ,  \frac{\delta S}{\delta g_{cd}( \vec x)}\right] - \alpha_A( \vec x) = 0.
\eeq
In 1968 \cite{Komar:1968aa} Komar took the  crucial step of establishing the parallel to non-singular Hamilton-Jacobi theory. He showed that there existed vanishing invariant phase space functions $\bar \beta^A \left[\vec x;g_{ab}(\vec x),p^{cd}(\vec x)\right] = \bar \beta^A_0 \left[g_{ab}(\vec x),p^{cd}(\vec x)\right] - \beta(\vec x) = 0$  canonically conjugate to $\bar \alpha_A$, and in addition $\beta^A(\vec x)$ satisfied the conventional non-singular Hamilton-Jacobi relation $\beta^A(\vec x) = \frac{\delta S}{\delta \alpha_A(\vec x)}$. The resulting canonical Poisson bracket algebra was the sought for Lie algebra of the factor group, the spacetime coordinate transformation group having been factored out. Komar called this factor group the ``proper canonical group". 

Now all of the elements were in place to investigate whether classical solutions of Einstein's equations could be recovered in the conventional manner by using the relation $\beta^A(\vec x) = \frac{\delta S}{\delta \alpha_A(\vec x)}$. As Komar stated, this would be a precondition for recovering a semi-classical limit in a WKB approximation in which $S$ would play the role of a quantum phase - although he did not pursue the WKB method in this paper. There did remain, however, a major hurdle, even at the classical level. One must still find commuting invariants $\alpha^0_A$.

Classically, of course, additional coordinate information is required to recover explicit solutions of Einstein's equations from solutions of the four functional differential equations for $S[g_{ab}; \alpha_A]$, (\ref{hjeqs}). The procedure that would normally correspond to WKB would be to prescribe the numerical values  $\alpha_A(\vec x)$. There would be no restrictions on the values of $g_{ab}(\vec x)$, and the $p^{ab}(\vec x) = \frac{\delta S}{\delta g_{ab}(\vec x)}$ will satisfy the constraints ${\cal H}_\mu(g_{ab}(\vec x), p^{cd}(\vec x)) = 0$.  Given the initial $g_{ab}(\vec x)$ and $p^{ab}(\vec x)$ one would select values for the lapse $N(\vec x,t)$ and shift $N^a(\vec x,t)$ and then evolve in $t$ using the Einstein equations $G_{ab}(\vec x,t) = 0$ (in the vacuum case). This is related to the program formulated later by \cite{Gerlach:1969aa}, to be discussed in Section \ref{sandwich}, but with the crucial difference that Gerlach did not require his $\alpha^0_A$ to be four-dimensional diffeomorphism invariants. 

But Komar suggested an alternative quantum approach which would correspond more closely with the standard procedure for finding classical solutions of the equations of motion for non-singular systems.  It is in principle possible, as in the classical Hamilton-Jacobi approach, to start with values for  $\beta^A(\vec x)$ and $\alpha_B(\vec x)$. One would of course need to have in hand the diffeomorphism invariants $\alpha^0_A(g_{ab}(\vec x), p^{cd}(\vec x))$ employed in (\ref{alphaa}) while the $\beta^B(\vec x)$ would be arbitrarily chosen explicit functions of $\vec x$. Then one could in principle invert the relation $\beta^A(\vec x) = \frac{\delta  S[g_{ab}; \alpha_B]}{\delta \alpha_A(\vec x)}$ to solve for the initial $g_{ab}(\vec x)$. But since these are only $2\times \infty^3$ relations one must employ four coordinate conditions. Komar suggested as an example $g_{ab} = 0$ for $a\neq b$ and $|g_{ab}| = 1$. The independent components could be, for example, $g_{11}$ and $g_{22}$, represented collectively as $g_{AA}$, $A = 1,2$. This would deliver an invariant form for the metric, defining
$$
\gamma_A(\vec x) := g_{AA}\left[\vec x; \alpha_B(\vec x), \beta^C(\vec x)\right]
$$
The $\gamma_A$ are a commuting set and they commute with all the components $g_{ab}$. One can then define the canonical variables $\pi^B(\vec x)$ as functionals of the $ \alpha_B(\vec x), \beta^C(\vec x)$ with Poisson brackets $\left\{\gamma_A((\vec x)) , \pi^B((\vec x'))\right\}:=  \delta_A^B \delta^3(\vec x-\vec x')$ .  Komar then observed that the set of constant ``observables $\gamma_A(x^a)$ characterizes the equivalence class of those tensor fields $g_{ab}(x^c)$ which can be mapped into each other by an element of the space-time subgroup of the full canonical group, or, equivalently, the equivalence class of those spatial geometries which can be stacked within the same Ricci-flat Riemannian manifold." But ``knowing this equivalence class does not yet determine the four-dimensional manifold. These are evidently inequivalent stackings of the same spatial geometries. The complementary set of observables $\pi^A(\vec x)$ must characterize the inequivalent stackings." He continues with the remark that Wheeler had identified this notion of inequivalent stackings in an undated and unpublished Princeton University report - but he stresses that in his case the $\pi^A(\vec x)$ are observable constants of the motion. This is likely an allusion to Wheeler's discussion of the 'thin sandwich conjecture' I will take up in the next Section. 

The paper concludes with a straight-forward quantization proposal : interpret the $\gamma_A(\vec x)$ and $\pi^B(\vec x)$ as Hermitian operators satisfying the standard algebra. The 3-metric operators would then be expanded in terms of these operators, and Komar proposes that polynomial expansions up to and including order two should be expressed as Hermitian symmetrized products. It is to be noted that "time" has disappeared entirely in this proposed quantum theory, and this is certainly related to one of the difficulties of Komar's approach listed in his conclusions, namely, what is ``the relationship of our interpretation of the quantities $\gamma_A(\vec x)$, $\pi^B(\vec x)$ to attributes which can be measured at least in principle."

Returning to another of the questions posed in the introduction to this section regarding the meaning of the principle function $S$, Komar was able in 1970 \cite{Komar:1970aa} to deduce its relation to the gravitational action. It turns out that it is equal to the modified gravitational action published by Dirac in 1958 \cite{Dirac:1958aa}. In fact, in the vacuum case since ${}^4R = 0$, the Einstein action vanishes. It is the divergence terms that Dirac removed from this action that contribute. Komar was able to show that using the Dirac Lagrangian ${\cal L}_D$, the Hamilton principal function is $S(t) =\int^t dt' d^3x {\cal L}_D$. As an additional bonus one finds that the principal function must be homogeneous of degree one in $g_{ab}$. Recall that Komar had shown in detail in using the original canonical variables that the action changes its numerical value under time evolution. (See Appendix B for a discussion of this fact, illustrated in detail for the free relativistic particle.) Now, due to the homogeneity of the total Hamiltonian generator, the action does not change under the action of the full four-dimensional diffeomorphism group, 

Komar's final paper in this series on the gravitational Hamilton-Jacobi approach was published in 1971 \cite{Komar:1971aa}. He succeeded in finding a canonical transformation to new  spatial metric and momentum variables $\bar g_{ab}$ and $\bar p^{cd}$ such that the scalar constraint $\bar {\cal H}_0$ was rendered homogeneous of degree $1/2$ in $\bar p^{cd}$. The implications for the behavior of $S$ under arbitrary infinitesimal coordinate transformations $\delta x^\mu = \delta^\mu_a \xi^a - \frac{g^{\mu 0}}{(-g^{00})^{1/2}} \xi^0$ are profound. The generator of these transformations is $H(\xi) = \int d^3x \left(\bar{\cal H}_a \xi^a +\bar{\cal H}_0 \xi^0 \right)$. Therefore
$\delta S =\int  d^3\!x \, \frac{\delta S}{\delta g_{ab}(x)} \frac{\delta H}{\delta p^{ab}(x)} = d^3\!x\,  p^{ab}(x) \frac{\delta H}{\delta p^{ab}(x)} = 0$ since the individual terms are proportional to the vanishing constraints $\bar {\cal H}_\mu = 0$. Thus $S$ depends only on the diffeomorphism invariants, $S = S[\alpha_A, \beta^B]$. Komar did entertain the possibility that as a consequence of employing potentially coordinate-dependent coordinate conditions in solving $\beta_A = \frac{\delta S}{\delta \alpha_A}$ for $\bar g_{ab}$ an explicit dependence on coordinates could appear. However, Bergmann's earlier proof rules out this possibility. I should note, however, that if intrinsic coordinates were employed as a coordinate condition, they will appear explicitly in $S$. This is illustrated for the free relativistic particle in Appendix B.  It is surprising that Bergmann and Komar never pursued this approach in the context of their gravitational Hamilton-Jacobi theory. Two possibilities come to mind. As we shall see, Wheeler's 'thin sandwich theorem' appeared to conflict directly with the reduced phase approach, and they did turn their attention to Wheeler's claims. And perhaps as importantly, the precise nature of the full four-dimensional Hamiltonian diffeomorphism symmetry was still unclear.

\section{Geometrodynamics, the sandwich conjecture, and the Wheeler-DeWitt equation} \label{sandwich}

Geometrodynamics received this designation by Wheeler, first in a classical context in which he sought a unified approach in which particles and fields would appear as manifestations of geometry \cite{Misner:1957ab}. This was to be the focus of Misner's thesis until it was revealed that Rainich had already obtained many of Misner's results.\footnote{Dean Rickles and I learned this in an interview we conducted with Misner in 2011.} At Wheeler's suggestion Misner then wrote his thesis on a Feynman-inspired path integral approach to quantum gravity \cite{Misner:1957aa}. In this context, of course, the action served the role of quantum phase, and it was in this connection that  Wheeler in 1957  \cite{Wheeler:1957aa} began to explore the implications with regard to the dependence of quantum field fluctuations on the size of the region over which the fluctuations occurred. Although Wheeler did not publish work directly related to Hamilton-Jacobi approaches until 1962, he clearly had earlier begun inquiring about what gravitational variables might be suitable in fixing initial and final states in a path integral approach. His senior thesis student David Sharp made what he viewed as major strides in this direction. He shared this thesis with Art Komar\footnote{Komar completed his own doctoral thesis \cite{Komar:1956aa} under Wheeler's direction in 1956.}, and also asked Peter Bergmann for comments. As noted in (\ref{deltasadm}), the non-vanishing temporal boundary variation of the gravitational action about solutions of Einstein's equations depends on the variation of the spatial metric $g_{ab}$. But it was clear to Wheeler that variations resulting from a change in the spatial coordinates would be unphysical, and that the quantum phase should therefore depend on the 3-geometry which he represented as ${}^3{\cal G}$. The question posed to Sharp was then: given this 3-geometry on an initial hypersurface with coordinate time $t'$ and a final hypersurface with coordinate time $t''$, did there exist a unique 4-geometry in the spacetime bounded by these surfaces? One might expect, if covariance under 3-diffeos exhausted the gauge freedom, that the answer would be yes. But there is of course an additional gauge freedom, and the manner in which Wheeler will deal with it will represents an essential irreconcilable difference with Bergmann and Komar. The sandwich conjecture is simply inconsistent with the reduced phase space approach advocated by Bergmann and Komar. According to their view the sandwich conjecture could only hold for 3-metrics that lie within the same equivalence class.

Misner did acknowledge that ``Bergmann finds that to carry through canonical quantization it may be necessary to find `true observables' in general relativity and use them in place of more  familiar variables."\footnote{\cite{Misner:1957aa}, p. 499} Indeed, he did propose a measure in his path integral approach that was purportedly invariant under the full four-dimensional diffeomorphism group. He also argued that the Hamiltonian generator of evolution vanished with the result that the Schr\"odinger and Heisenberg pictures merged. But he did suppose that quantum states would be represented by functionals of ${}^3{\cal G}$, and it was not clear how this choice could be consistent with the necessity that physically measurable variables must be invariant under the full four-dimensional invariance group. It is noteworthy, however, that he did propose an `intrinsic' time - essentially the total 3-volume of a closed universe. But this volume is itself not invariant under the action of the full group.

Returning to Sharp, perhaps since he was an undergraduate beginner with no historical baggage, his idea was to consider a so-called `thin sandwich' and return to the original Lagrangian approach to general relativity in which the initial value problem contemplates fixing on an initial spacelike hypersurface both the 3-metric and the time derivative of the 3-metric. There is of course already in this view an assumption concerning the choice of time coordinate. The thin sandwich notion arises from the view that this specification can be thought of as prescribing ${}^3{\cal G}$ on an initial $t =$ constant surface and on an infinitesimally neighboring surface. It should not be a surprise that the choice of local time inherent in a selection of time derivatives carries implications on the choice of lapse and shift functions on the initial time hypersurface, and this is what Sharp set about finding in expressing the ADM action in terms of $g_{ab}$, $\dot g_{cd}$, $N_a,$, and $N$.
In detail, the ADM Lagrangian, as given in the 1962 paper by Baierlain, Sharp, and Wheeler \cite{Baierlain:1962aa} takes the form
\beq
{\cal L}_{ADM} = \left( {}^3g\right)^{1/2}\left[ N {}^3\!R + \frac{1}{4 N} \left( \left(\kappa^a_a\right)^2 - \kappa_{ab} \kappa^{ab}\right) \right],
\eeq
where $\kappa_{ab} := \dot g_{ab} - N_{a|b} - N_{b|a}$ and indices are raised with the inverse of the 3-metric, $e^{ab}$ and the `$|$` denotes covariant differentiation with respect to the 3-metric. Varying with respect to $N$ yields
\beq
N = \pm \frac{1}{2} \left( \kappa^a{}_b \kappa_a{}^b - \kappa^2 \right)\left({}^3\!R\right)^{-1/2}.
\eeq
This lapse can then be substituted back into the action. Varying the result with respect to $N_a$ yields second order differential equations for $N_a$. The explicit expressions are not given in this article. 

In a letter from Wheeler to Bergmann, dated May 5, 1960, Wheeler says that with respect to the neighboring constant time hypersurfaces $\sigma'$ and $\sigma''$ `` not only can ${}^3\!{\cal G}'$ and ${}^3\!{\cal G}''$ be freely given, without supplementary condition, but \underline{also}, one thereby \underline{automatically} fixes the separation of $\sigma'$ and $\sigma''$ (through the field equations, on which one calls to produce a $V_4$ compatible with ${}^3\!{\cal G}'$ and ${}^3\!{\cal G}''$). (Analogy: In mechanics, give $x'$ and $t'$, and $x''$ and $t''$, and give $\ddot x + \omega^2 x = 0$; then the history is determined (except for exceptional cases); in gen. rel., ${}^3\!{\cal G}'$ is analogous to the \underline{combination} of $x'$ and $t'$; ${}^3\!{\cal G}''$ to the \underline{combination} of $x''$ and $t''$. The separation does not have to be \underline{given}; it is automatically implied). As Sharp is only a senior, but very promising, I do hope he gets credit for this point that seems to me a very new \& very important question of principle. His thesis has been typed and it is to be multilithed \& you are to get a copy."\footnote{Syracuse University Bergmann Archive (SUBA), correspondence folder.} Wheeler wrote this in response to a query of Bergmann's in a detailed four page commentary he made on Wheeler's draft copy of Geometrodynamics and the Problem of Motion.\footnote{The final version was published in 1961 \cite{Wheeler:1961aa}.} Bergmann wrote ``I am intrigued by the statement in the table for which you quote Sharp's thesis: that knowledge of the configuration variables on two hypersurfaces is sufficient to fix the field between. This is obviously true in a Lorentz-covariant theory, where the distance between two surfaces is given by the differences in the coordinates. It is considerably less obvious in a general-relativistic theory (your third column), where knowledge of the coordinate values tells us next to nothing about the actual locations of the two hypersurfaces relative to each other. Further, in the same table: If you adopt as your Hamiltonian data $g_{ik}$, $K_{ik}$. there are four restrictions on these twelve variables (Dirac's $H_L$, $H_s$), whereas you seem to list only one, in addition to the two electromagnetic restrictions. Or do I misunderstand the next-to-last line of entries?"\footnote{SUBA, correspondence folder.} (Apparently what Wheeler refers to as the the  'Poynting flux ", namely the constraints $H_s = 0$,  conditions had been missing in this draft.) In a follow-up letter from Bergmann to Wheeler, (with a copy to Komar) dated July 3, 1960, Bergmann acknowledged having received the Sharp thesis from Komar, and he raises several questions. He brings to Wheeler's attention a corresponding analysis performed in 1959 by Peres and Rosen \cite{Peres:1959ab}. They had obtained the same differential equations for the lapse, working directly with the four vacuum Einstein equations $G^{\mu 0} = 0$. They are of course simply the constraints ${\cal H}_\mu = 0$, expressed in terms of $g_{ab}$, $\dot g_{cd}$, $N_a,$, and $N$. They then investigated the linearized version of these equations. The major question, raised also in the cited paper, concerns whether the differential equations for the shift are well posed, and if so whether the solutions are unique. Bergmann writes `` Sharp, as you know, claims that he has proven that in the three-dimensional hypersurface the Cauchy-Kowalewski problem is correctly set. Thus he essentially states that if we give the $g_{0k}$ freely (subject to inequalities only) on a two-dimensional spacelike surface, then in a three-dimensional vicinity of this surface the $g_{0k}$, $g_{00}$ are uniquely determined, and with them the $K_{ij}$."\footnote{SUBA, correspondence folder.} He points out that in the linearized theory the $G^{0m}$ constraints give an inhomogeneous differential equation for the $N_a$ of the form $\vec \nabla \times \vec \nabla \times \vec N = \vec S$, with the result that the divergence of $\vec S$ must vanish. "It is therefore pertinent to search whether in the full theory also the Bianchi identities may not bring it about that the four constraint equations satisfy among themselves an identity that makes the construction of the Cauchy-Kowalewski problem impossible."\footnote{SUBA, correspondence folder.} 

Bergmann then began to address some of the complications that arise in the full theory, citing the form (\ref{shift}) exhibited by Peres and Rosen,
in a particularly compact form in 1959 \cite{Peres:1959ab}, which using the present notation, is
\beq
e^{ab} \left[\left(N^{-1} \kappa_{ac}\right)_{|b} - \left(N^{-1} \kappa_{ab}\right)_{|c}\right] = 0. \label{shift}
\eeq
Bergmann rewrote this as\footnote{Bergmann wrote $N$ instead of $N^{-1}$, but this did not affect his argument.}
\beq
0=\left[N^{-1} \left(\kappa^b{}_c - \delta^b_c\kappa \right)\right]_{|b}.
\eeq
 ``If you substitute for [$\kappa_{ab}$] the full expression, it turns out that this equation contains only anti-symmetric derivatives of [$\vec N$] in the square brackets on the extreme right. I lack the time to see what happens after $N$ has been substituted, but I am made wary by the realization that such a substitution is not likely to render the Bianchi identity (the one with the index zero) impotent." He goes on to write ``Even if it could be shown that the full theory has a decent Cauchy-K problem on the three-dimensional hypersurface, I still cannot follow Sharp's conclusion that in the four-dimensional domain the $g_{mn}$, $\dot g_{mn}$ represent a satisfactory set of Cauchy data. Clearly, in order to fix the $g_{00}$, $g_{0k}$, he must (if we are to assume that my earlier doubts have been resolved in favor of Sharp's paper) choose the $g_{0k}$ arbitrarily on one 2-surface. Can we believe that this choice does not affect the intrinsic geometry at all, in other words, that the $K_{ij}$ will come out more or less the same regardless of the choice? Remember, the ultimate claim is that giving the $g_{mn}$, $\dot g_{mn}$ on an arbitrary 3-surface \underline{determines} the 4-geometry in the vicinity of the 3-surface. How can we be assured that the additional choice of the $g_{ok}$ on a 2-surface fixes only coordinates but has no effect on the geometry of the four-dimensional manifold? Merely counting degrees of freedom will not help. Essentially, we should have to show, either that the $K_{ij}$ are unaffected by the choice on the 2-face, or that the change in $K_{ij}$ corresponds to a coordinate transformation on the 3-face + deformation of the 3-face (change of imbedment) such that the $g_{ij}$ remain completely unchanged and the $K_{ij}$ change exactly as required by a change in the choice of the $g_{0k}$ on the 2-face. The latter sounds formidable, but the decision can be reached in finite time, and without the expenditure of ingenuity, by considering the effect of \underline{infinitesimal} changes in the $g_{ok}$ on the 2-face. Then, given an initial solution of the whole problem, the propagation of these changes and their conversion into changes of the $K_{ij}$ becomes a linear problem."\footnote{SUBA, correspondence folder.}
 
Wheeler responded on July 6, 1960. ``Your \underline{most} stimulating and helpful letter about Sharp's thesis came yesterday; and tonight we have been working through the paper of Rosen \& Peres, which we had not seen before." ``Of course they did not investigate the elliptic character of the equation, as you point out; but on this Sharp has done much more, in showing what goes on as to the coefficients of the second derivatives. Nevertheless, he points out that the usual classification does not apply when the coefficients of the 2nd derivatives depend, as here, on the first derivatives. That is why he limited his statements, as he does, to the determinant of the second derivatives in any \underline{local} region; no classification as to being `elliptic' or `hyperbolic' appears appropriate over \underline{extended} regions, under these circumstances, except of course with respect to solutions that differ only infinitesimally from a given solution. We expect to discuss your points more and undoubtedly Sharp or I will write you further. The result that only ${}^3\!g'_{mn}$ and ${}^3\!g''_{mn}$ have to be given is so important that, as you point out, no loose ends should be left in the analysis."\footnote{SUBA, correspondence folder.}

 Komar also shared his copy of the Sharp thesis with Misner, who then communicated directly with Sharp (with copies to Bergmann and Wheeler) on July 7, 1960.\footnote{SUBA, correspondence folder.} Misner referred to Bergmann's July 3 letter. He did agree that some qualms about the thin sandwich in the full theory were legitimate, but he did disagree with the Peres and Rosen claim that the program did not work in the linearized theory. He did agree with Peres and Rosen that in the linearized theory the lapse is undetermined. In addition, only the transverse lapse field $\vec N_T$ is determined, with the freedom in the longitudinal field $\vec N_L$ reflecting a gauge freedom that manifests itself as a coordinate transformation of the extrinsic curvature. There is also a restriction on $\dot g_{ab}$, and Misner points out that this restriction is consistent with the proven constancy, with appropriate asymptotic boundary conditions, of the total energy in the full theory. This constancy was proven by Arnowitt, Deser, and Misner earlier in 1960 \cite{Arnowitt:1960ae}.
 
Bergmann discussed these points with Komar in a letter dated July 11, 1960, writing approvingly of Misner's treatment of the linearized case. But he writes ``There remains one serious hair in the ointment: The differential equations obeyed by the [$\vec N$] is of the form curl curl $\vec N = \vec \sigma$, where div $\vec \sigma = 0$ is guaranteed by the fourth constraint equation. With the standard boundary conditions at spatial infinity this differential equation does indeed fix [$\vec N_T$], provided we look for global solutions, free of singularities and of multiple connections. As we know, for the full theory the existence of analogous theorems is neither known nor too likely; Sharp, of course, was careful not to examine the spatially global problem, but to confine himself to the local problem in three, and in four, dimensions." ``To the extent that one may place any faith into the heuristic value of the linearized problem, I should say that $g_{mn}$, $\dot g_{mn}$ (subject to the fourth constraint , as Charlie pointed out) will be a satisfactory set of Cauchy data globally, in that they fix the identity of the Riemann-Einstein manifold; they are not satisfactory locally, in that \underline{in}equivalent R-E manifold patches will exhibit the same initial data."\footnote{SUBA, correspondence folder.}

There is no mention of this debate in \cite{Baierlain:1962aa}, other than an observation that appropriate boundary conditions on the lapse ``are essential". 
Wheeler did expand further on the thin sandwich conjecture in his 1962 Warsaw lectures \cite{Wheeler:1964aa}. Here he clearly indicates the quantum mechanical source of his attachment to the sandwich conjecture, ``No one has found any way to escape the conclusion that geometrodynamics, like particle dynamics, has a quantum character. Therefore, the quantum propagator, not the classical history, is the quantity that must be well-defined", and this propagator in the Feynman path approach proposed by Misner depends on the temporally initial and final 3-geometries. The motivation is spelled out even more explicitly in his 1963 Les Houches lectures \cite{Wheeler:1964ab}:``Sharp's two-surface formulation of dynamics has been found (1) to provide an evident point of contact between the propagator theory  of quantum theory and Hamilton's principal function of classical theory and (2) to give a clear picture of what quantities in a system (a) can be freely specified and (b) can thereby be used to pick out a dynamical history." In Warsaw  Wheeler offered a new interpretation of Mach's principle, namely that it was a statement that solutions of Einstein's equations were to be excluded if they did not satisfy chosen boundary conditions. In this context the ambiguity in the solutions for the shift vector argued that the universe must be spatially closed! He writes ``It would be an enormous labor to take up one by one all the questions that are left unanswered here and treat them systematically. Moreover, there is wanting one key element in the discussion - proof that the solution of the variational problem [varying $N^a$ in the action] (when there is a solution) is unique." There is no reference to the issues that had been broached by Bergmann in the correspondence cited above. In a follow-up article in 1964 \cite{Wheeler:1964ad} with the same title Wheeler does briefly examine the uniqueness question. He says ``Unfortunately the three coupled second-order equations to be solved are only quasi-linear, not linear. The problem appears difficult without resorting to deeper mathematical considerations which are not immediately apparent. Therefore no decisive results can be offered here." He then goes on to consider the linearized version for the spatially closed universe, concluding that ``It is conjectured that there is no other independent solution [other then those he displays] which are free of truly geometrical singularity, as distinguished from coordinate singularity, over the entire three-sphere." One finds a similar disclaimer in \cite{Wheeler:1964ab} in a 
section entitled ``A Key Principle Today can be Formulated only as a Conjecture."

Both Bergmann and Komar did follow up with their own analyses of the conjecture. In 1970 \cite{Bergmann:1970aa} cites Komar's yet unpublished analysis in which he employs the Hamiltonian expression for the time derivative of the 3-metric,
\beq
\dot g_{mn} = - \frac{1}{2} \left(N_{m|n} +N_{n|m} \right) + \frac{N}{\sqrt{{}^3\!g}} \left( p_{mn} - \frac{1}{2} g_{mn} p^s{}_s\right), \label{dotg}
\eeq
to solve for $p^{mn}$ and then substitutes into the constraints. The scalar constraint, when ${}^3\!R \ne 0$, yields an algebraic solution for $N$, and this in turn can be substituted into $p^{mn}$, with $p^{mn}{}_{|n}$ finally yielding a system of second-order partial differential equations for the three variables $N^m$ that is inhomogeneous-linear in $N^m$.  Bergmann then asked what would be the consequences if solutions of these equations existed. Would there exist unique neighboring solutions which would then satisfy simpler linear inhomogeneous equations. He concluded that ``nothing general can be said about the elliptic, hyperbolic, or parabolic character of the system of equations."\footnote{ \cite{Bergmann:1970aa}, pp. 51-52.} 
Komar \cite{Komar:1970ac} himself wrote out the full explicit set of equations that were to be satisfied by the lapse functions - excluding the case where ${}^3\!R$ vanishes. He concluded that a local spacelike region solutions could be found ``which are neither unique or stable".\footnote{\cite{Komar:1970ac}, p. 821.} He noted in addition that in the linearized problem there are thin sandwich assignments which lead to mutually incompatible equations for the lapse and shift. One year later Komar \cite{Komar:1971ac} proposed a generalization of the conjecture involving a canonical transformation to a new three-metric for which the conformal factor involved the momentum. He was able to prove that the resulting differential equations were elliptic. On the other hand, in transforming back to the original canonical variables he encountered a more severe restrictive condition on the value of the curvature scalar ${}^3\!R$.

Wheeler was undeterred in his embrace of the sandwich conjecture as central to the `plan of general relativity'. And he maintained this stance in spite of demonstrations by his own students and conversation partners that the conjecture had limited validity. I refer in particular to his senior thesis student Ohanian \cite{Belasco:1969aa}. Christodoulou and Francaviglia
 \cite{Christodoulou:1979aa} employed geometrical arguments to show that ``the thin-sandwich conjecture is false "\footnote{\cite{Christodoulou:1979aa}, p. 480.}, and they they acknowledge discussion with Wheeler.\footnote{The definitive status of the conjecture seems to be that enunciated by Bartnick and Francaviglia \cite{Bartnik:1993aa}, namely that the conjecture is valid provided certain geometrical conditions are assumed, and they are ``not generically satisfied". This result was generalized by Giulini \cite{Giulini:1999aa} to include specific material sources.}

Bryce DeWitt already in 1960 held views that were similar to Wheeler's in regard to the role to be played by 3-geometries in an eventual quantum theory of gravity. In his 1960 essay submitted to the Gravitational Research Foundation he refers to Higg's who he claims ``was the first to show explicitly that the secondary constraints of the Dirac theory are the generators of infinitesimal coordinate transformations", but Higgs dealt only with spatial coordinate transformations. The lesson for Higg's was that the quantum wave function could not depend on the lapse, and that $\Psi({}^3\!g_{mn})$ must be invariant under spatial diffeomorphisms. Referring to Dirac, and the work of Arnowitt, Deser, and Misner, DeWitt observes that ``A byproduct of this activity has been the revelation that the Hamiltonian formalism, with the preferred status it gives to the time coordinate ... is in large measure a theory of the embedding of a positive definite 3-space ($t$ = constant) in a 4-space of signature $\pm$ 2." This is consistent with Wheeler's view that four-dimensional covariance has been lost - to be replaced by what he calls ``multi-fingered time". This notion first appears in print in his Battelle lectures \cite{Wheeler:1968aa}: ``One has long known that time in general relativity is a many-fingered entity. The hypersurface drawn through spacetime to give one ${}^3\!{\cal G}$ can be pushed forward in time a little here or a little there or a little somewhere else to give one or another or another new ${}^3\!{\cal G}$. `Time' conceived in these terms {\it means} nothing more or less than {\it the location of the ${}^3\!{\cal G}$ in ${}^4\!{\cal G}$}. In this sense `3-geometry is the carrier of information about time' ." The idea is depicted diagrammatically for the first time in Figure 15-1 of \cite{Wheeler:1964aa} showing the now familiar honeycomb structure illustrating the meaning of the lapse and shift functions. This is or course a natural outgrowth of the sandwich conjecture - and a rejection of the primacy of four-dimensional diffeomorphism covariance. 

Wheeler did not bring the Hamilton-Jacobi approach to bear in print in his approach to quantum gravity until his 1967 Battelle lectures. We have a clear indication of the impetus for this new departure from Bryce DeWitt \cite{DeWitt:1999aa}. He explained at the Eighth Marcel Grossmann Meeting in 1997 ``John Wheeler, the {\it perpetuum mobile} of physicists, called me one day in the early sixties. I was then at the University of North Carolina in Chapel Hill, and he told me that he would be at the Rayleigh-Durham airport for two hours between planes. He asked if I could meet him there and spend a while talking quantum gravity. John was pestering everyone at the time with the question: What are the properties of the quantum mechanical state functional $\Psi$ and what is its domain? He had fixed in his mind that the domain must be the space of 3-geometries, and he was seeking a dynamical law for $\Psi$. I had recently read a paper by Asher Peres which cast Einstein's theory into Hamilton-Jacobi form, the Hamilton-Jacobi function being a functional of 3-geometries. It was not difficult to follow the path already blazed by Schr\o dinger and write down a corresponding wave equation. This I showed to Wheeler, as well as an inner product based on the Wronskian for the functional differential wave operator. Wheeler got tremendously excited at this and began to lecture about it on every occasion. I wrote a paper on it in 1965, which didn't get published until 1967 \cite{DeWitt:1967aa} because my Air Force contract was terminated, and the Physical Review was holding up publication of papers whose authors couldn't pay the page charges. My heart wasn't really in it because, using a new kind of bracket discovered by Peierls, I had found that I could completely dispense with the cumbersome paraphernalia of constrained Hamiltonian systems ..." I have not been able to establish the precise date of this meeting. There is no mention of this new approach in his application to the Office of Naval Research for support that was to commence in December, 1964\footnote{Bryce DeWitt Papers (BDP), Briscoe Center for American History}, but the report on results of research conducted under that grant does contain the statement 
 ``One of the senior investigators (B. S. D.) has recently proposed a functional  equation for the state functional of the gravitational field in the so-called canonical theory. He has proposed a functional integral expression for the inner product of two state vectors. This work has not yet been published but it has stimulated considerable activity both at North Carolina and at Princeton and has made it possible for the first time to discuss a number of fundamental issues in concrete form. Chief among these is the problem of gravitational collapse."
 In addition, ``In a reappraisal of the canonical quantum theory of gravity the attempt has been made to find a dynamical interpretation of the so-called `fourth Hamiltonian constraint'. This has led to the discovery of a six-dimensionlonal differential hyperbolic manifold which underlies the intrinsic dynamics of the gravitational field and the introduction of a functional differential `wave' equation of the second degree for the state functional of the theory. This in turn has led to the discovery of a natural definition for the inner product of two state vectors."\footnote{BDP} DeWitt's work was split into three parts at the request of the editor.\footnote{ \cite{Dewitt-Morette:2011aa}, p 60} The first in the trilogy is devoted to DeWitt's introduction in the canonical approach of a replacement of the canonical momentum by the operator $- i\frac{\delta}{\delta g_{ij}}$. The constraint ${\cal H}_i = 0$ then yields the condition that the wave function can depend only on the 3-geometry ${}^3\!{\cal G}$. The scalar constraint then yields what has become known as the Wheeler-DeWitt (WD) equation,
 \beq
G_{ijkl} \left( \frac{\delta}{\delta g_{ij}} \frac{\delta}{\delta g_{kl} }+ \left({}^3\!g\right)^{1/2} {}^3\!R \right) \Psi\left[{}^3\!{\cal G}\right] = 0, \label{WD}
 \eeq
 where $G_{ijkl} := \frac{1}{2}  \left({}^3\!g\right)^{-1/2} \left(g_{ik} g_{jl} +g_{il} g_{jk} - g_{ij} g_{kl} \right)$ is regarded as a metric on a six-dimensional manifold $M$ with hyperbolic signature $-+++++$. A dilatation of $g_{ij}$, defined as $\zeta := (32/3)^{1/2}\left({}^3\!g\right)^{1/4} $ corresponds to a `timelike' displacement. The form of (\ref{WD}) suggested to DeWitt a natural 'Klein-Gordon' inner product
 \beq
 \left(\Psi_a, \Psi_b\right) := Z \int_\Sigma \Psi^*_b \left[{}^3\!{\cal G}\right] \times \Pi_x \left(d\Sigma^{ij} G_{ijkl} \frac{\overrightarrow{ \delta}}{i \delta g_{kl}} -  \frac{\overleftarrow{\delta}}{i \delta g_{kl}}G_{ijkl} d\Sigma^{ij} \right) \Psi^*_a \left[{}^3\!{\cal G}\right]. 
 \eeq
 As DeWitt observed, this product suffers an analogous `negativie probabilty' interpretational problem because of the appearance of the second functional derivative with respect to the `time' coordinate. DeWitt also pointed to `certain suggestive features' of the WD equation, consistent with Wheeler's sandwich conjecture. ``A question now arises as to the extent the Riemannian structure of $M$ may be regarded as imposing a structure on ${\cal M}$ [the set of all 3-geometries] by way of the Hamiltonian constraints ... The existence of a timelike coordinate in $M$ suggests that a corresponding `intrinsic time' exists in ${\cal M}$ and that the Hamiltonian constraint does indeed have a dynamical content." ``If we regard the usual enumeration of the degrees of freedom possessed by the gravitational field, namely {\it two} for every point of 3-space, as being valid in a finite world, this leaves one quantity per 3-space to play the role of intrinsic time. Baierlein, Sharp, and Wheeler have shown in the classical theory that if the intrinsic geometry is given on any two hypersurfaces then, except in certain singular cases, the geometry of the entire space-time manifold, {\it and hence the absolute time lapse between the two hypersurfaces}, is determined. Moreover, it is determined by the constraints. Analogously, the quantum theory is completely determined by the transformation functional $\left< {}^3\!{\cal G}'\left.\right| {}^3\!{\cal G}'' \right>$, where $\left.\left.\right| {}^3\!{\cal G} \right>$ denotes that state of the gravitational field for which there exists at least one hypersurface having an infinitely precise geometry ${}^3\!{\cal G}$. Wheeler has emphasized the importance of the two-hypersurface formulation of gravidynamics (or `geometrodynamics' as he calls it) and has suggested the use of the Feynman sum-over-histories method to compute the transformation functional." DeWitt also considered the semiclassical limit corresponding in analogy with Schr\o dinger's limit, to a wave function of the form $\Psi\left[{}^3\!{\cal G}\right] = {\cal A} \exp (i {\cal W})$ where $ {\cal A}$ and $ {\cal W}$ are assumed to be real functionals satisfying the restriction $\left| \frac{\delta  {\cal A}}{\delta g_{ij}}\right| <<  \left| {\cal A}\frac{\delta  {\cal W}}{\delta g_{ij}}\right|$, and supposing that $ {\cal W}$ satisfies the Hamilton-Jacobi equation
 \beq
 G_{ijkl}  \frac{\delta  {\cal W}}{\delta g_{ij} } \frac{\delta  {\cal W}}{\delta g_{kl} } =\left( {}^3\!g\right)^{1/2} {}^3\!R. \label{EHJ}
 \eeq
 DeWitt observes that a unique 4-geometry can be computed from a given solution of (\ref{EHJ}) by making the identification $\pi^{ij} = \frac{\delta {\cal W}}{\delta g_{ij}}$ and then integrating the Hamiltonian equation
 \beq
 \frac{\partial g_{ij}}{\partial x^0} = 2 N G_{ijkl} \frac{\delta {\cal W}}{\delta g_{kl}} + N_{i,j} + N_{j,i}, \label{dewitt}
 \eeq
 where the lapse and shift are arbitrary other than the requirement that $N>0$. DeWitt refers in this regard to the as yet unpublished work by Gerlach \cite{Gerlach:1969aa}, an abstract of his results having been published the previous year \cite{Gerlach:1966aa}. 
 
Gerlach made use of the `constants of integration' that appear in solutions of the Hamilton-Jacobi equation. He argued that there were two fields which he called $\alpha(x^i)$ and $\beta(x^i)$ that could then be used to characterize the complete solution so that the complete ${\cal W}$ was ${\cal W}({}^3\!{\cal G}; \alpha, \beta)$. The contrast with the Komar approach is significant. Firstly,  the number of independent configuration variables according to Komar should be two - as he recognized in his imposition of coordinate conditions. This should translate into a recognition that the Hamilton-Jacobi principal function should depend on $2 \times \infty^3$ metric variables, and  not $3 \times \infty^3$. The difference is due to the fact that Gerlach assumes that it is  ${}^3\!{\cal G}$ that is prescribed, and not invariant under the action of the four-dimensoional diffeomorphism group. In addition, as Gerlach does note but without analyzing its significance, since there are only four functional differential equations for ${\cal W}$ yet six independent momenta $\pi^{ij}$, there remains a freedom in their specification through the relation $\pi^{ij} = \frac{\delta {\cal W} }{\delta g_{ij}}$. Recall that Komar eliminated this freedom through the use of his four-diffeomorphism invariants $\alpha^0_A[g_{ij}, \pi^{kl}] = \alpha_A(x^m)$, with two corresponding functional differential equations (\ref{alphaa}) to be satisfied by ${\cal W}$. Since these conditions are not imposed by Gerlach it is clear that his $\alpha(x^i)$ and $\beta(x^i)$ cannot be interpreted as diffeomorphism invariants. Furthermore, they cannot be interpreted as identifying equivalence classes because the additional freedom in the choices of $g_{ij}$ and $\pi^{kl}$ permits the transition to new equivalence classes without altering $\alpha(x^i)$ and $\beta(x^i)$. This is of course consistent with the DeWitt-Gerlach-Wheeler view that the  elements of superspace are 3-geometries which are not invariant under diffeomorphisms that alter coordinates in the direction perpendicular to spatial hypersurfaces.
As summarized by by Misner, Thorne, and Wheeler \cite{Misner:1973aa}, the conditions that delivers classical solutions are $\frac{\delta {\cal W}}{\delta \alpha} = \frac{\delta {\cal W}}{\delta \beta} = 0$. And as noted previously by DeWitt, quantum superpositions of solutions of the Wheeler-DeWitt equation will in the limit of small $\hbar$ rely on the satisfaction of the Einstein Hamilton-Jacobi equation (\ref{EHJ}) to deliver, with the input of lapse and shift as in (\ref{dewitt}), wave packets that satisfy Einstein's equations.

Wheeler embraced the DeWitt-Gerlach Hamilton-Jacobi view in his 1967 Battelle Lectures \cite{Wheeler:1968aa}, and it became a centerpiece of his program in quantum geometrodynamics. He assigned the name `superspace' to the space whose points represented specific closed 3-geometries, being orbits of $g_{ij}$ under the action of the spatial diffeomorphism group. Fischer initiated a careful study of this space \cite{Fischer:1970aa}, concluding that it was not a manifold, but was partitioned into ``manifolds of geometries, the strata, such that the geometries of high symmetry are completely contained in the boundary of geometries of lower symmetry." Differing geometries could have differing topologies, and Wheeler contemplated also two distinct spin structures associated with wormholes. DeWitt later proposed an 'extended superspace' which did possess a manifold structure\cite{DeWitt:1970aa}. 

It is a curious fact that Wheeler never abandoned the sandwich conjecture, even though it is clear that wave packets obtained through the procedure described above will never describe evolution from one equivalence class to another - as is clear from the classical evolution prescription (\ref{dewitt}). There are abundant different 3-geometries that can never evolve into each other. 
And this is inconsistent with Misner's formulation of the Feynman path integral. There will exist many transition amplitudes $\left<{}^3\!{\cal G'}\left.\right|{}^3\!{\cal G}''\right>$ that vanish.

This situation apparently contrasts with the Bergmann-Komar Hamilton-Jacobi approach. Recall that for Komar the `constants' $\alpha_A(x^a)$ are the numerical values of actual constants of the motion, i.e., invariants under the action of the full four-dimensional diffeomorphism group. And in addition, the two additional functional differential equations (\ref{alphaa}) satisfied by $S$ require that these invariants take these values. The result is that Komar's $\alpha_A$ or conjugate $\beta^B$ really do label equivalence classes of metrics, i.e., metrics that cannot be transformed into each other through a general coordinate transformation. In this framework the transition function for the 3-metrics on to given spatial hypersurfaces vanishes if the metrics do not lie in the same equivalence class.  Komar offered in 1971 another related criticism of the quantum geometrodynamics program. ``The intractability of the fourth constraint has led a number of authors [citing Wheeler \cite{Wheeler:1964ab} and DeWitt \cite{DeWitt:1967aa}] to propose that in the quantization program one should treat this constraint differently from the first three. In particular, these authors suggest that instead of eliminating the fourth constraint, thereby obtaining observables, one should convert the corresponding Hamilton-Jacobi equation into the analogous Schr\o dinger equation via substituting [into the scalar constraint] the operator relation $p^{mn} = -i \frac{\delta}{\delta g_{mn}}$. Two criticisms which have been lodged at such an approach are (i) the questionability of replacing quantities which are not space-time observables (constants of the motion) by Hermitian operators, and (ii) the lack of symmetry in the treatment of the four constraints which generate the full coordinate-invariance group of the Einstein theory. A correlated and more immediate criticism is that, should we for the moment envision having obtained a solution of the Schr\o dinger equation and compute the expectation value of $p^{ij}$ and $g_{ij}$ for the state determined by this solution, $\left<p^{ij}\right>$ and $\left<g_{ij}\right>$, respectively, the classical functions so obtained would not satisfy the fourth constraint equation (since it is nonlinear). Thus the expectation values of the canonical variables so computed would not satisfy an Ehrenfest theorem, and wave packets would not determine a mean Riemann-Einstein manifold. If however, we employ observables, such a difficulty can never arise."\footnote{\cite{Komar:1971aa}, p. 924}

I conclude this section with a few observations on the nature of the interactions between Bergmann and Wheeler and their associates. I have heard from several participants in the regular monthly meetings that were organized by Jim Anderson and held at the Stevens Institute of Technology in Hoboken, New Jersey, from the late 1950's to the mid 1970's, that the substantial interaction was always courteous and respectful.\footnote{My direct sources here include Jim Anderson, Dieter Brill, Joshua Goldberg, Karel Kuchar, Charles Misner, Ted Newman, Ralph Schiller, and Louis Witten.} Nevertheless, there are indications of an underlying dynamic that could have played perhaps an unconscious  role in the participants' responsiveness to ideas presented at these sessions. Ted Newman notes that that he felt there was a 'mild hostility' between Peter and Wheeler.\footnote{Private communication}. Karel Kuchar speaks of a ``subdued tension between Bergmann and Wheeler, but nothing that would really come out as a clash".\footnote{Interview that I and Charles Torre conducted with Kuchar in April, 2016.} It was his impression that Bergmann thought of himself as a ``natural successor" to Einstein. But Wheeler, while ``a very modest man when it came to evaluating himself", justifiably judged that he himself had achieved a great deal in relativity. This impression of Bergmann is consistent also with a comment from Stanley Deser, that Bergmann behaved as though he ``owned the field".\footnote{Interview that Dean Rickles and I conducted with Deser in March, 2011.}  I might mention here also two remarkable incidents that were witnessed by Steven Christensen.\footnote{private communication.}. He reported that in Copenhagen in 1971 he overheard Bergmann saying negative things in the lounge about Bryce DeWitt, after DeWitt had presented a talk on 
``The Theory of Superspace". In addition he witnessed a ``stunning" confrontation in Trieste later that year in which DeWitt accused Bergmann of ``not giving credit to others that Bryce believed were being ignored by Bergmann". I must add that as student of Bergmann in Syracuse in the mid-1970's, it is true that we did hear criticisms of particle physics inspired, non-geometric approaches to quantum gravity, but I never heard personal attacks. On the contrary, Bergmann embodied for me the exemplar of a gentleman of science. 

\section{Conclusions}

The focus in this article is on the extension of Hamilton-Jacobi techniques to general relativity. It was appropriate to begin with the Poincar\'e-Cartan form and its intimate relation to end point variations about solutions of classical dynamical equations. These relations were continued to field theory by Weiss, and subsequently by Schwinger. But difficulties arose in dealing with gauge theories. As had been recognized by both, the Hamilton principal function needed to depend of gauge invariant fields. This discussion established the background for work begun in the late 1960's by Peter Bergmann and Arthur Komar.

I have mainly traced in this review their efforts  to identify aspects of the underlying general covariance of Einstein's theory that in their view should survive in some form in an eventual quantum theory of gravity. It was clear from the start that spacetime coordinates could not play a fundamental role in the ultimate theory. Their hope and expectation was that only by focusing on physically observable quantities in the classical theory could one gather insights into the foundations of this ultimate theory. Thus from the beginnings in 1949 the emphasis was on the group of local symmetry transformations. These transformations were studied in both Hamiltonian and Lagrangian classical formalisms, with the intention of identifying true physical variables that did not change their values under the action of the symmetry group. And most importantly they wished to discover the algebra which these objects must satisfy, in other words, to find the transformations engendered by the corresponding Hermitian operators. Formulated in the phase space of general relativity, the objective was to identify as physical the orbits described under the action of the symmetry group - each orbit representing a physical observable. In other words, the objective was to determine the structure of the reduced phase space of the theory. There was no doubt from the beginning in the `Syracuse school' that there was some sense in which the full four-dimensional diffeomeorphism group was realized in this classical phase space. And it was insistence on this point that in the 1960's led to a divergence in approach between the Syracuse school and quantum geometrodynamics approach of Bryce DeWitt and John Archibald Wheeler. Although each promoted a Hamilton-Jacobi perspective they disagreed on the underlying fundamental symmetries. Although Wheeler and DeWitt did of course impose the scalar constraint, they did not recognize its role as a symmetry generator. Rather, they interpreted this constraint as merely engendering an advancement in what Wheeler dubbed multifingered time. Consequently, since it was no longer viewed as a generator of a gauge symmetry, objects that were to represent physical observables were not required to remain invariant under its action. As we have seen in the discussion of gauge symmetry in general in the context of Hamilton-Jacobi theory and its use in a WKB approximation to quantum theory, in particular in regard to electromagnetism, one could sensibly define transition amplitudes only for gauge invariant quantum variables. Since Wheeler did not recognize this symmetry it was natural for him to consider transition amplitudes for variables that were invariant under the restricted spatial coordinate transformation group, namely 3-geometries ${}^3\!{\cal G}$.

Bergmann and Komar, on the other hand, put their stress on 4-geometries. We have seen that Bergmann, his students, and his associates investigated various methods for constructing true diffeomorphism invariants that would ultimately play the role of quantum gravitational observables. Many earlier methods involved the imposition of gauge conditions and related constructions of `extended Dirac brackets' (which I have called Bergmann-Goldberg brackets).Ultimately, the resulting reduced phase space was coordinatized with the diffeomorphism-invariant (constants of the motion) $\alpha^0_A$ and $\beta_0^{B}$, leading to what they called a `frozen formalism'. Bergmann noted in 1977 \cite{Bergmann:1977ab} that its adoption  ``is unpalatable to many, as it appears to eliminate from the formalism all semblance of dynamical development." But he then hinted at a procedure which could be employed to recover dynamics - by transforming to new canonical coordinates. In fact, the suggestion was that this could be done in the context of the Hamilton-Jacobi approach.

This actually suggests another line of reasoning originally due to Komar, namely the introduction of intrinsic coordinates based in the vacuum case on the use of Weyl curvature scalars. It is surprising that neither Bergmann nor Komar investigated the use of these intrinsic coordinates in the context of the Hamilton-Jacobi approach. This is the main subject of the following paper \cite{Salisbury:2021ad}.\footnote{An initial discussion appears in \cite{Salisbury:2016ab,Salisbury:2018aa}, with further corrections in https://arxiv.org/abs/1508.01277v6.} It leads to a formalism based on the reduced algebra in which variables undergo non-trivial evolution in intrinsic time. A choice of spacetime intrinsic coordinates and of an $\alpha^0_A\left(g_{ab},p^{cd}\right)$ does fix a representative solution of Einstein's equations, including the lapse and shift as in the thin sandwich conjecture. This has profound implications for an eventual quantum theory of gravity.

\section*{Acknowledgements}
Thanks to Alex Blum for his criticaI reading of an earlier incomplete draft of this paper. And I would like to thank J\"urgen Renn and the Max Planck Institute for the History of Science for support offered me as a Visiting Scholar.

\section*{Appendix A. Generalizations of the Hamilton-Jacobi Methods}

De Donder, Weyl, and Carath\'eodory established the foundations for a generalization of the Hamilton-Jacobi equations, and I will in particular briefly describe the Carath\'eodory theory as nicely summarized by von Rieth \cite{Rieth:1984aa}.\footnote{This is also essentially the prepublication summary that Carath\'eodory communicated to Weyl in 1935 in response to Weyl's article. It is to be found on page 210 in the excellent  Carath\'eodory  biography by Maria Georgiadou \cite{Georgiadou:2004aa} . }  It is basically an extension of the Poincar\'e-Cartan form $p_i \delta q^i - H(q,p) \delta t$  to field theory. We wish to minimize the action $A = \int d^4\!x {\cal L}(x^\mu,\phi_A(x), \partial_\mu \phi_A(x)$. There exists an equivalent action obtained by subtracting a total divergence $\Phi(x, \phi, \partial \phi)= \frac{dS^\mu(x,\phi)}{dx^\mu}=\partial_\mu S^\mu(x,\phi) + \frac{\partial S^\mu(x,\phi)}{\partial \phi_A} \partial_\mu \phi_A $ from this Lagrangian. The fundamental idea is to place restrictions on unknown functions $S^\mu(x,\phi)$ and $\psi_{A\mu}(x,\phi)$  with $\bar{\cal L}(x,\phi,\partial \phi) := {\cal L}(x,\phi,\partial \phi) - \Phi(x, \phi,\partial \phi)$, such that
 \beq
 \bar{\cal L}(x,\phi,\psi) = {\cal L}(x,\phi,\psi)  - \partial_\mu S^\mu(x,\phi) - \frac{\partial S^\mu(x,\phi)}{\partial \phi_A} \psi_{ A\mu}(x,\phi)= 0 \label{zeroLhat}
 \eeq
and $\bar{\cal L}(x,\phi,v(x,\phi)) > 0$ when $v_{A \mu}(x,\phi) \ne \psi_{A\mu}(x,\phi)$. Since by these assumptions the variation of $\hat{\cal L}(x,\phi,v)$ about $v_{A \mu}(x,\phi) = \psi_{A\mu}(x,\phi)$ vanishes, we have when varying only $v_{A\mu}$, $0 = \frac{\partial {\cal L}(x,\phi,v)}{\partial v_{A\mu}}\left.\right|_{v = \psi} \delta v_{A\mu}- \frac{\partial S^\mu(x,\phi)}{\partial \phi_A} \delta v_{A\mu}$. Therefore we conclude that 
\beq
\left.\frac{\partial {\cal L}(x,\phi,v)}{\partial v_{A\mu}}\right|_{v = \psi} =  \frac{\partial S^\mu(x,\phi)}{\partial \phi_A} \label{SA}
\eeq
 and (\ref{zeroLhat}) becomes
\beq
\frac{\partial {\cal L}(x,\phi,\psi)}{\partial \psi_{A\mu}} \psi_{A\mu}(x,\phi) = {\cal L}(x,\phi,\psi)  - \partial_\mu S^\mu(x,\phi) \label{partialL}
\eeq

Finally we can derive Carath\'eodory's first main result: Consider a new action
\beq
\hat A := \int d^4\!x \left[ {\cal L}(x,\phi,\psi)  + \frac{\partial {\cal L}(x,\phi,\psi)}{\partial \psi_{A\mu}}\left(\partial_\mu \phi_A - \psi_{A \mu} \right)  \right].
\eeq
Substituting from (\ref{partialL}) and employing  (\ref{SA}) this becomes
\beq
\hat A = \int d^4\!x \left( \partial_\mu S^\mu(x,\phi)  + \frac{\partial S^\mu(x,\phi)}{\partial \phi_A}  \partial_\mu \phi_A \right) = \int d^4\!x \frac{d S^\mu}{dx^\mu} = \oint S^\mu d \Sigma_\mu,
\eeq
where in the last step we apply Stoke's theorem. Thus when we assume that the values of $\phi_A$ are fixed on the 3-dimensional closed spacetime boundary the integral is independent of $\phi_A$ in the interior! This is an extension of David Hilbert's  ``independence theorem"  to which Weiss' thesis advisor Born referred in his  1934 \cite{Born:1934ab} article.\footnote{The reference is actually to Hilbert's G\"ottingen lectures on Variationsrechnung which Born attended in 1914. The original published report of the ``Unabh\"angigkeitssatz" appeared in 1900 \cite{Hilbert:1900aa}, with greater detail in 1905 \cite{hilbert:1905aa}.} As observed by Born, in a model in which there is only one independent variable $t$ and $L$ has no explicit $t$ dependence we have
\beq
\hat A =  dt \left( L(\phi(t),\psi(t,\phi(t)) + \frac{\partial L(\phi(t),\psi (t,\phi(t))}{\partial \psi_A} \left(\dot \phi_A(t) -\psi_A(t,\phi(t))  \right)\right) =:   p^A d\phi_A - H dt, \label{dA}
\eeq
where $p^A(\phi,\psi) := \frac{\partial L(\phi,\psi)}{\partial \psi_A}$ and $H(\phi,\psi) := -L(\phi,\psi)+ \frac{\partial L(\phi,\psi)}{\partial \psi_A} \psi_A $. The fact that (\ref{dA}) is an exact differential leads to a set of differential equations that must be satisfied by the unknown ``geodesic fields" $\psi_A(\phi)$. Because $\frac{\partial \hat A}{\partial \phi_A} = p^A$ and $\frac{\partial \hat A}{\partial t} = -H$ we have as a consequence that $\frac{\partial p^A}{\partial \phi_B} - \frac{\partial p^B}{\partial \phi_A} = 0$ and $\frac{\partial H}{\partial \phi_A} - \frac{\partial p^A}{\partial t} = 0$. These are differential equations to be satisfied by the $\psi_A(t,\phi)$. But as Born points out the most efficient way to solve them is to solve first for $\hat A$ by replacing the $p^A$ argument in $H$ by $\frac{\partial \hat A}{\partial \phi_A}$ - where it is assumed, of course, that one is dealing with a non-singular theory for which one can solve for the velocities $\psi_A$ in terms of the momenta $p^B$. The result is the standard Hamilton-Jacobi equation
$$
\frac{\partial \hat A}{\partial t} + H\left(\phi_B, \frac{\partial \hat A}{\partial \phi_A} \right) = 0.
$$

Carath\'eodory and de Donder did extend these methods to field theory, but as far as I can tell their treatments did not include models in which arbitrary gauge symmetries were present. In the non-singular case where one can solve for $\psi_{A\mu}$ in terms of the generalized canonical momenta $p^{A\mu} := \frac{\partial {\cal L}(x,\phi,\psi)}{\partial \psi_{A\mu}}$, and ${\cal H} := p^{A\mu} \psi_{A\mu} - {\cal L} = {\cal H}(x,\phi,p)$. One then obtains generalized canonical equations of the form $\partial_\mu \phi_A = \frac{\partial {\cal H}}{\partial p^{A\mu}}$ and $\partial_\mu p^{A\mu} = - \frac{\partial {\cal H}}{\partial \phi_A}$. And according to (\ref{partialL}) we obtain the generalization of the Hamilton-Jacobi equation\footnote{de Donder called this a ``g\'en\'eralisation du th\'eor\`eme direct de Jacobi", \cite{Donder:1935aa}, p 116}
\beq
\partial_\mu S^\mu(x,\phi) + {\cal H}\left( x,\phi,\frac{\partial S^\mu(x,\phi)}{\partial \phi_A}\right) = 0. \label{genHJ}
\eeq

There is a mystery regarding de Donder's incorporation of a special kind of covariance symmetry in his formalism. He did consider so-called parameterized finite dimensional and field theoretic models. He showed that a necessary and sufficient condition for reparameterization covariance was that the Lagrangian be homogeneous of degree one in derivatives with respect to the parameters. The corresponding generalized Hamiltonian was therefore of degree zero, with the consequence that what we now call a constraint would arise. One particularly relevant finite dimensional case is the relativistic free particle, and in this case the analysis is equivalent to that first developed by Rosenfeld in 1930 \cite{Rosenfeld:1930aa,Rosenfeld:2017ab},\footnote{See \cite{Salisbury:2017aa} for an analysis of this work.} and further amplified by Bergmann and Dirac. The puzzle for me is that he did succeed in deriving the generalized Hamiltonian for Einsteinien gravity with the corresponding generalized Hamilton and Hamilton-Jacobi equations, yet he did not address the underlying general covariance.\footnote{An initial analysis parameterized generally covariant theory was first undertaken by Bergmann \cite{Bergmann:1949aa} and Bergmann and  Brunings \cite{Bergmann:1949ab}. The mystery deepens with the recognition that Rosenfeld had collaborated with de Donder in Paris in 1927, and had returned to Paris in 1931 to give a four-week course at the Institut Poincar\'e  (see \cite{Bustamante:1997aa}, p. 58) following the completion his groundbreaking paper on constrained Hamiltonian dynamics. Rosenfeld would then presumably have discussed his work with de Donder before the latter published his book on the calculus of variations.}

\section*{Appendix B. The free relativistic particle}

I will illustrate the Bergmann Komar Hamilton-Jacobi analysis with the free relativistic particle with spacetime position $q^\mu$ as a function of the paramter $\theta$. Consider the reparameterization covariant Lagrangian $L_p = - m \left(\dot -q^2\right)^{1/2}$, where $\dot q^\mu := \frac{dq^\mu(\theta)}{d \theta}$.  We have $p^\mu = m \dot q^\mu/\left(-\dot q^2\right)^{1/2}$ and the Hamiltonian constraint $H = p^2 + m^2 = 0$, with the Hamilton-Jacobi equation $\eta^{\mu \nu}\frac{\partial S}{\partial q^\mu} \frac{\partial S}{\partial q^\nu} + m^2 = 0$. First I address Bergmann's 1966 argument that $S$ cannot depend on $\theta$. It follows here from the identity
\beq
0 \equiv \frac{\partial H\left(q,\frac{\partial S(q,\theta)}{\partial q}\right)}{\partial q^\mu} = \left. \frac{\partial H\left(q,\frac{\partial S(q,\theta)}{\partial q}\right)}{\partial q^\mu}\right|_{p} + \frac{\partial H}{\partial p^\nu}\frac{\partial^2 S}{\partial q^\nu \partial q^\mu} = \frac{\partial H}{\partial p^\nu}\frac{\partial^2 S}{\partial q^\nu \partial q^\mu}.
\eeq
On the other hand,
\beq
\dot p_\mu = 0 = \frac{d}{d \theta} \frac{\partial S}{\partial q^\mu} = \frac{\partial H}{\partial p^\nu}\frac{\partial^2 S}{\partial q^\nu \partial q^\mu} + \frac{\partial^2 S}{\partial q^\mu \partial \theta},
\eeq
so we conclude that $ \frac{\partial^2 S}{\partial q^\mu \partial \theta} = 0$, and therefore $\frac{\partial S}{\partial \theta} = f(\theta)$, which does not affect the dynamics and can be taken to be zero.

Next I consider Komar's selection of invariants in identifying equivalence classes of solutions. This is really a selection of gauge conditions, in a manner that will be discussed in detail in \cite{Salisbury:2021ad}. As Komar notes, without these conditions it is not possible to solve the Hamilton-Jacobi equation for the momenta in terms of the configuration variables. The simple choice I will make here is 
\beq
\bar \alpha^a = p^a - \alpha^a = 0,
\eeq
 where the $\alpha^a$ are numerical constants, but there are of course many other possibilities.  The $p^a$ is the analogue of $\alpha^0_A$ in (\ref{alphaa}). These are vanishing invariants under evolution in $\theta$ since as we know $\dot p^a = 0$. Thus we now have a set of four Hamilton-Jacobi equations which are
\beq
\frac{\partial S}{\partial q^a} = \alpha^a,
\eeq
and consequently, from $H\left(p\right) = 0$,
\beq
\frac{\partial S}{\partial q^0} = - \left(\vec \alpha^2 + m^2 \right)^{1/2}.
\eeq
We conclude that $S = \alpha_a q^a - \left(\vec \alpha^2 + m^2 \right)^{1/2} q^0$.

Continuing with the extension of Komar's analysis to this model, we have
\beq
\beta^a  = \frac{\partial S}{\partial \alpha} = q^a - \frac{\alpha^a}{\left(\vec \alpha^2 + m^2 \right)^{1/2} } q^0 = q^a - \frac{p^a}{\left(\vec p^2 + m^2 \right)^{1/2} } q^0. \label{beta}
\eeq
This is indeed invariant under infinitesimal reparamterizations $\theta' = \theta - \epsilon(\theta)$ on the constraint surface $p^0 =\left(\vec p^2 + m^2 \right)^{1/2}$,   whereby $\delta q^\mu = \dot q^\mu \epsilon$, so
\beq
\delta \left(q^a - \frac{\alpha^a}{\left(\vec \alpha^2 + m^2 \right)^{1/2} } q^0\right) = \epsilon (-\dot q)^{1/2} \left(p^a -\frac{p^a}{\left(\vec p^2 + m^2 \right)^{1/2} } p^0\right)  = 0.
\eeq
Continuing with the notation used in representing the numerical value of the invariant, I will represent this numerical value by $\beta^a$ and the corresponding phase space function by $\beta_0^a := q^a - \frac{\alpha^a}{\left(\vec \alpha^2 + m^2 \right)^{1/2} } q^0$. Then we have the vanishing invariant $\bar \beta^a = \beta_0^a  - \beta^a = 0$.
Note also, as in Komar's vacuum gravitational case, the invariant phase space functions satisfy the canonical Poisson bracket $\left\{\bar \beta_a, \bar \alpha^b \right\} = \delta^b_a$.

Of course, from (\ref{beta}) we obtain the free relativistic particle solution in the chosen gauge, namely $\theta = q^0$,
\beq
q^a = \beta + \frac{\partial f}{\partial \alpha^a} + \frac{p^a}{\left(\vec p^2 + m^2 \right)^{1/2} } q^0. \label{qa}
\eeq

Notice also that just as was noted by Komar in the case of general relativity, the action $S$ actually undergoes a change under the action of the reparameterization group,
\beq
\delta S = p_a \delta q^a - \left(\vec p^2 + m^2 \right)^{1/2} \delta q^0 = -\epsilon \left(- \dot q^2 \right)^{-1/2}m^2
\eeq
 Komar connected this change with a change under time evolution. One must however distingusih between evolution and transformations of solution trajectories under diffeomorphisms. This is spelled out in detail in \cite{Pons:1997aa} and will be discussed further in \cite{Salisbury:2021ad}. A specific field dependence is required to obtain variations in configuration-velocity space that are projectable under the Legendre map to phase space. In this model the infinitesimal descriptor of projectable reparameterizations take the form $\epsilon =  \left(- \dot q^2 \right)^{-1/2} \xi(\theta)$. And under these reparamterizations the action undergo the $q^0 = \theta$ dependent variations $-\xi(q^0)m^2$.  

I observe finally that the numerical value of  $p^a =\alpha^a  $ fixes an equivalence class under reparameterizations. The invariant $\bar \beta^a$ alters this value.

\bibliographystyle{apalike}
\bibliography{qgrav-V19}

\end{document}